\documentclass[fleqn,usenatbib]{mnras}

\usepackage{newtxtext,newtxmath}

\usepackage[T1]{fontenc}
\usepackage{ae,aecompl}

\usepackage{graphicx}	
\usepackage{amsmath}	
\usepackage{epsfig}
\usepackage{epstopdf}

\def\be{\begin{equation}}
\def\ee{\end{equation}}

\title[Initial modification,  TBTF \& void phenomenon]{ Addressing the too-big-to-fail problem and the void phenomenon through a modified initial power spectrum}
\author[Hamed Kameli, Shant Baghram]{
Hamed Kameli $^{1,2}$
~and Shant Baghram $^{1,2}$ \thanks{baghram@sharif.edu}
\\
$^{1}$
Department of Physics, Sharif University of
Technology, P.~O.~Box 11155-9161, Tehran, Iran\\\
$^2$Research Center for High Energy Physics, Department of Physics, Sharif University of Technology, Tehran 11155-9161, Iran
}
\date{Accepted XXX. Received YYY; in original form ZZZ}
\pubyear{2025}
\begin{document}
\label{firstpage}
\pagerange{\pageref{firstpage}--\pageref{lastpage}}
\maketitle
\begin{abstract}
We investigate the impact of early-time initial conditions on nonlinear structure formation and evolution within the framework of the semi-analytical Excursion Set Theory (EST). Our analysis reveals that adding a Gaussian bump to the initial curvature power spectrum at small scales enhances the abundance of massive halos while sharply reducing the number of small-mass halos, and consequently, satellite galaxies. Moreover, this modification increases the frequency of major mergers while suppressing high-mass-ratio minor mergers. These features may offer resolutions to the missing satellite and Too Big to Fail (TBTF) problems.
In underdense regions -voids- the same modifications increase the likelihood of finding massive halos embedded in voids while similarly decreasing the small-halo population, and consequently, faint galaxies. This behavior suggests a potential solution to the void phenomenon, in which embedded halos, despite being too massive, were too rare to be noticed.
More precisely, our results indicate that an excess of massive structures emerges at mass scales near the center of the Gaussian bump: $k_* = 1.85 \,\rm{h/Mpc}$ and $k_* = 3.95 \,\rm{h/Mpc}$. These scales correspond to mass scales of $M_* = 10^{11}$ and $M_* = 10^{10}$, respectively. This modification extends up to two orders of magnitude in higher mass scales, while reducing the abundance of halos below $M_*$ by two to three orders of magnitude. Additionally, we find that evolutionary conditions, halo-in-halo, and particularly halo-in-void statistics serve as more sensitive and complementary probes for differentiating among cosmological models. 
\end{abstract}




\section{Introduction}
\label{Sec:1}
Cosmology is standing on the brink of change. On one hand, its standard model, cosmological constant plus cold dark matter, known as $\Lambda$CDM has a great achievement in describing the cosmic microwave background (CMB) data \citep{Aghanim:2018eyx} and Large-scale structure (LSS) surveys \citep{Alam:2016hwk}. On the other hand, it confronts many tensions and challenges.
The recent Dark Energy Spectroscopic Instrument (DESI) DR2 data \citep{2025JCAP...02..021A}, which refer to a dynamical dark energy model and a decade-long $H_0$ tension \citep{{2021CQGra..38o3001D}} discrepancy between observation of local standard candle \citep{2019ApJ87685R} and the CMB data \citep{Aghanim:2018eyx}, are examples challenging the standard picture of the accelerated Universe. Also, we should mention challenges of cold dark matter (CDM) in galactic scales, such as missing satellite problems \citep{Klypin:1999uc,Moore:1999nt} and too big to fail problems \citep{BoylanKolchin:2011de,2015PNAS..11212249W,Bullock:2017xww}. The missing satellite problem refer to the observations, where we find an almost one order of magnitude less satellite galaxies around the Milky way in comparison to the prediction of N-body simulation in standard model. The Too Big to Fail (TBTF) problem refer to satellite massive halos that are too big (with deep gravitational potential) to fail to host the baryonic counterpart.

Furthermore, these challenges are shown in under-dense regions, such as void phenomena. This phenomena is related to lack of expected faint galaxies within voids. In other words, the voids are emptier than simulation prediction in standard cosmological model \citep{peebles:2001void,Tinker:2009void}, and the distribution and statistics of galaxies in voids \citep{2021ApJ...916L..24T}. The Large Scale Structure (LSS) is a fertile and promising probe or ruler to investigate alternate cosmological models and investigation of these small scale challenges and tensions \citep{kameli:2022Hten, Parkavousi2023void}.

 Many large-scale structure surveys are working now and collecting huge data sets \citep{2016arXiv161100036D,EUCLID:2011zbd} and also a couple of them are on their way \citep{2020PASA...37....2W,2022arXiv220307220B}. Accordingly, modelling and observation of the formation and evolution of structures have great importance. 
In this direction, we pursue studies by investigating the effect of initial conditions on late time formation and statistics of structures (for example see \cite{2016JCAP...11..014C,Fard:2017oex}).
In this work we study the deviation from standard initial condition and its effect of the statistics of halos, voids, halo-in-halo and halo-in-voids statistics.

The core idea of this work is using the LSS as a probe to find possible deviations from standard model of cosmology. This idea studied in literature vastly    \citep{Elgaroy:2001wu,Baghram:2014nha,Namjoo:2014nra,Hassani:2015zat,Fard:2017oex}. For modifications in initial condition see \cite{Garrison-Kimmel:2014kia,Leo:2017wxg}.
For works involving the initial condition impacts on LSS see \cite{hahn2011multi,agarwal2014cons} and, specifically for initial non-Gaussianity and their impact on LSS see \cite{slosar2008cons,desjacques2010pri,tseliakhovich2010nona}. Also, for  suppression of matter power spectrum in small scales see \cite{Nakama:2017ohe}.  

In this work, we propose a toy model by adding a Gaussian bump to the initial curvature power spectrum in small scales. This modification affect the large scale structure formation and evolution, and their statistical properties. These enhancement and modification in initial condition could be created in the context of inflationary models and primordial features.
For pioneering works on inflationary models which altered the initial power spectrum, see \cite{Salopek:1988qh,starobinsky:1992spe,Randall:1995dj,Adams:1997de,Martin:2000xs,Hunt:2004vt}. 
In this work, following our previous investigation \citep{kameli:2020mod}, we investigate the effect of modified initial conditions on missing satellite problems, Too Big to Fail (TBTF) problems and void phenomena.

 In our previous work, we studied the effect of modified initial condition on Dark Matter (DM) halos. Here, we study its mutual effects on void and halo-in-void statistics. This study serve as a complementary investigation on the effect of initial conditions on voids and the structures embedded in void. For previous studies see \cite{Sheth:2004hie,paranjape:2012hie,Parkavousi2023void,Jennings2013void,DAmico2011void,kameli:2022Hten}. \\
We use the context of Excursion set theory (
EST) \citep{Bond:1990iw, Zentner:2006vw,Nikakhtar:2016bju}.
As more realistic and accurate EST model, we investigate the statistics of the structure with non-Markov extension of the EST as well. Different non-Markov trajectory approaches are developed in the literature. \cite{maggiore2010non,Ma2011non} have used a perturbative schemes, which derived by a path integral formulation and an alternative approach is investigated by \cite{Musso:2014non1,Musso:2014non2}. Alternatively, we used the Cholesky decomposition approach for generating non-Markov trajectories proposed by \cite{Nikakhtar:2018qqg} and its extension for realistic cosmological model \citep{kameli:2020mod}. We also incorporate the ellipsoidal collapse in our EST model which used a scale dependent barrier \citep{2001MNRAS.323....1S}.\\
{{We should emphasise that recent simulations show that supernova-driven outflows, reionization heating, and	disk-shocking can lower central densities and eliminate TBTF in many halos. The gravitational potential from the central galaxy causes enhanced tidal stripping in satellites that is not present in dark-matter-only simulations, making subhalos more susceptible to mass loss and enhancing disruption of dwarf galaxies \citep{2016MNRAS.457.1931S}. In this work, we suggest that with our non-baryonic power-spectrum approach, we can address the small-scale challenges of CDM \citep{2022NatAs...6..897S}. Some promising avenues could be in this direction, such as (i) coupling the baryonic feedback prescriptions to EST barriers and (ii) using empirical halo-occupation models to connect halos to galaxies.  }} \\
{{We also should note that the recent developments in the small-scale challenges of DM such as  Gaia data, particularly from Gaia DR3, has provided valuable insights into this problem, both by revealing new satellite galaxies and by highlighting areas of incompleteness in current observations \citep{2025A&A...696L..19T}. Gaia data has led to the discovery of numerous new dwarf galaxies and satellite galaxy candidates, helping to bridge the gap between simulations and observations \citep{2019A&A...623A.129F}. Also, there are studies to relax the tension by considering the complexities of baryonic physics and its relation with DM \citep{2018PhRvL.121u1302K}. One step further, new studies with a semi-analytic model of galaxy formation tend to solve the missing satellite problem and even predict more satellites than the prediction of $\Lambda$CDM \citep{{2025MNRAS.540.1107S}}. An odd that finds its counterpart in observation, such as the recent report by the Hyper Suprime-Cam (HSC) Subaru Strategic Program (SSP) survey team \citep{2024PASJ...76..733H}. We will also discuss how our idea can address these new findings.
}} \\
The structure of this work is as follows: Section \ref{Sec:2} devoted to the theoretical background. We will discuss the structure formation from linear to non-linear regime in Excursion Set Theory context. We would present the analytical calculation of different large scale structure statistics. Section \ref{Sec:3}, we represent our results, for the impact of a modified initial curvature power spectrum on the large scale statistics and discussed how this modification can resolve the TBTF, missing satellite and void phenomenon simultaneously. Finally in section \ref{Sec:5}, we have our conclusion and future remarks.

\section{ Structure formation from linear to non-linear regime}
\label{Sec:2}

\begin{figure}
	\includegraphics[width=\columnwidth]{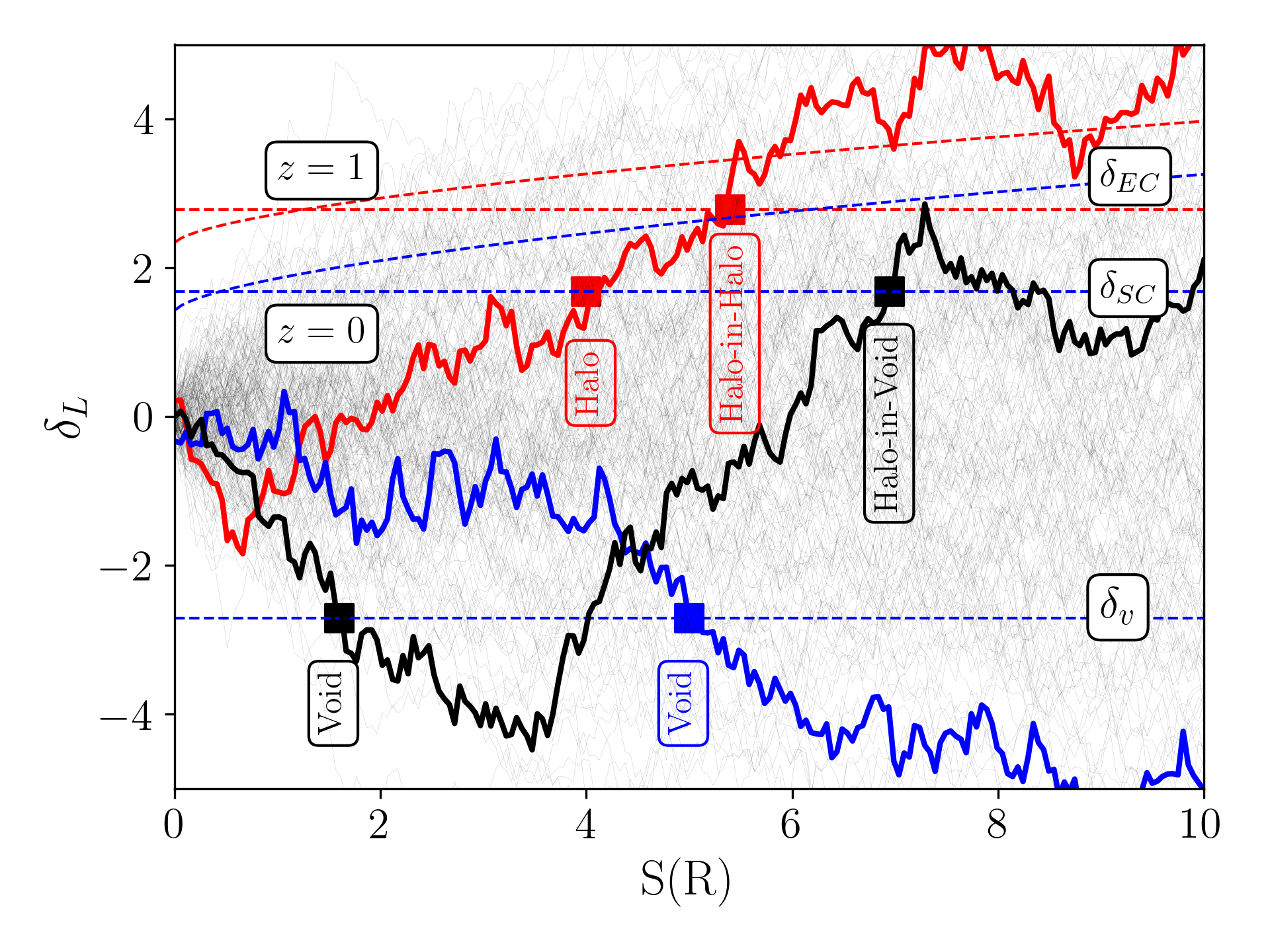}
	\caption{This figure provides a pedagogical illustration of barrier crossing and computational approaches within the framework of Excursion Set Theory (EST). The blue-dashed lines denote the constant Spherical Collapse barrier ($\delta_{\rm{SC}}$) and the variance-dependent Ellipsoidal Collapse barrier ($\delta_{\rm{EC}}$) at $z=0$, while the red-dashed lines depict the corresponding barriers at $z=1$. A large ensemble of Markov trajectories is displayed in the background, illustrating the stochastic nature of trajectories and structure formation. The red trajectory highlights the first up-crossing event associated with halo formation at $z=0$ and a second crossing for the halo-in-halo scenario at $z=1$. In the EST framework, halos form when trajectories first up-cross the collapse barrier and the halo-in-halo phenomena observed by a conditional barrier crossing. The blue trajectory demonstrates the formation of voids, characterized by the first crossing of the negative void barrier ($\delta_v$). Lastly, the black trajectory exemplifies the halo-in-void scenario, wherein the trajectory initially crosses the void barrier then up-crossing the collapse threshold. This correspond to forming halos embedded in voids. All crossing points are marked with bold squares to emphasize critical transition events.		
} \label{fig:traj}
\end{figure}

The standard paradigm of structure formation is based on a hierarchical framework, wherein larger and more massive gravitationally bound objects emerge from the mergers of smaller structures. This process is primarily driven by gravitational instability and the nonlinear growth of initial matter perturbations. An old yet sophisticated formalism was introduced in the pioneering work of Press and Schechter (PS)  \citep{Press:1973iz}. The fraction of dark matter halos forming at late times with a specific halo mass $M$ is corresponded to the probability distribution function of the matter density contrast, $\delta_{\rm{M}}$, exceeding the critical density threshold $\delta_c$. The critical overdensity $\delta_c$, in a smoothed region, is extrapolated linearly to the present epoch as
\begin{equation}
	f(\delta_{\rm{M}}>\delta_c)=\int_{\delta_c}^{\infty} d\delta_{\rm{M}}\frac{1}{\sqrt{2\pi}\sigma(M)}\exp\left[-\frac{1}{2}\frac{\delta_{\rm{M}}^2}{\sigma^2(M)}\right],
\end{equation} 
where $\sigma(M)$ is the variance of perturbations in the comoving scale of $R$ related to the mass of DM halo $R=(3M/4\pi\rho_{\rm{M}})^{1/3}$. Hence, the variance is considered as a function of mass or radius $\sigma(R)$ interchangeably. Linearly evolved density perturbations are small so the variance can be calculated using the linear matter power spectrum with a proper window function (smoothing function) as
\begin{equation} \label{variance}
	S\equiv\sigma^2(R)=\frac{1}{2\pi^2}\int dk k^2 P_{m}(k,z=0)\tilde{W}^2(k;R),
\end{equation}  
where $P_{m}(k,z=0)$ is the linear matter power spectrum at present time. $\tilde{W}(k;R)$ is Fourier transform of smoothing window function.
The matter power spectrum in a linear regime is formulated as follows:
\begin{equation}
	P_m(k,z)=A_{l}k^{n_s}T^2(k)D^2(z),
\end{equation}
where $D(z)=\delta_{\rm{M}}(k,z) / \delta_{\rm{M}}(k,z =0) $ is the scale dependent growth function normalized to unity at the present epoch, $T(k)$ is the Transfer function, $A_{l}$ is the amplitude of perturbation in late time and $n_s$ is the spectral index.

The fundamental equation that connects late-time structure formation to initial conditions and the gravitational potential is the Poisson equation. The gauge-invariant gravitational potential, $\Phi(k,z)$, is related to the density contrast via $k^2 \Phi(k,z) = 4\pi G \rho_{\rm{M}}(z) (1+z)^{-2}\delta_{\rm{M}}(k,z)$. The evolution of $\Phi(k,z)$ across different cosmological epochs is determined by the Einstein equations. The gravitational potential at a given redshift is expressed in terms of its initial value as $\Phi(k,z) = \frac{9}{10}T(k) D(z) (1+z) \Phi_{\rm{ini}}$, where $T(k)$ is the transfer function (We use the Eisenstein-Hu transfer function \citep{Eisenstein:1997ik}), and $\Phi_{\rm{ini}}$ represents the initial potential, which is linked to the curvature perturbation ${\cal{R}}_k$ as $\Phi_{\rm{ini}} = \frac{2}{3} {\cal{R}}_k$.
The curvature perturbation in the standard cosmological model with nearly scale-invariant initial conditions, is parameterized as
\be
{\cal{P}}_{\cal{R}} (k) = \frac{1}{2\pi^2}k^3 P_{\cal{R}}(k)=A_s (\frac{k}{k_p})^{n_s -1},
\ee
where $P_{\cal{R}}\equiv |{\cal{R}}_k|^2$,  $A_s$ represent the amplitude of primordial power spectrum, $n_s$ is the spectral index of perturbations and $k_p $ denote the pivot wavenumber, using  the values reported by Planck 2018 \citep{Aghanim:2018eyx}.
The relation between the matter power spectrum and the curvature power spectrum is given by
\be
	P_m(k,z) = \frac{8\pi^2}{25}\frac{k}{\Omega_M H_0^4} T^2(k)D^2(z) {\cal{P}}_{\cal{R}} (k).
\ee \\
The number density of DM-halos can be deduced from PS formalism. They argue that the fraction of DM-halos with mass greater than $M$ is proportional to $f(\delta_M>\delta_c)$ as  $F(>M)=2f(\delta_M>\delta_c)$. The adapted fudge factor $2$ is a hint of not counting the negative density perturbations on a large scale \citep{2012MNRAS.420.1429P}.
An extension to a more sophisticated version of this framework is known as the Excursion Set Theory. The number density of dark matter halos is given by
\begin{equation}
	n(M,t)dM=\frac{\bar{\rho}}{M}\frac{\partial F(>M)}{\partial M}dM,
\end{equation}
where $\bar{\rho}$ is the background density. It is convenient to define the height parameter.  $\nu=\delta_c(z)/\sigma(M)$. This parameter allows the number density to be expressed in terms of a universal function as $f_{\rm{PS}}$:
\begin{equation}
	n(M,t)dM=\frac{\rho}{M^2}f_{\rm{PS}}(\nu)|\frac{d\ln\nu}{d\ln M}|dM,
\end{equation}
where $f_{\rm{PS}}$ is PS universality function:
\begin{equation}
	f_{\rm{PS}}(\nu)=\sqrt{\frac{2}{\pi}}\nu \exp[-\nu^2/2].
\end{equation}

Various modifications have been introduced to this function to incorporate the physics of DM-halo formation and collapse models. Revisiting the fudge factor of 2, it hints at a more sophisticated approach to structure formation process, linking the linear regime to the nonlinear evolution. The framework of Excursion Set Theory (EST) is designed to map the statistical properties of linearly evolved overdense regions to the population of DM-halos. In a two-dimensional space defined by the linearly evolved density contrast $\delta_L$ versus the variance $S$, trajectories are plotted to represent the stochastic behavior of density contrast across different smoothing scales.  
 
Fig. \ref{fig:traj} illustrates this concept, where trajectories start form large scales (low variances) and walk toward smaller scales (higher variances). This characteristic is unique feature of the hierarchical structure formation, where variance depends monotonically on scale and decreases as a function of mass (smoothing scale). The density contrast $\delta_L$ evolves linearly from its initial value via the growth function. In the EST context, a DM-halo forms when a trajectory first up-cross the critical density barrier. For spherical collapse, this barrier is given by $\delta_{\rm{SC}}\simeq 1.69$. In the conventional formulation of EST, trajectories are computed at the present time. To examine structure formation probabilities at higher redshifts, the critical density barrier is adjusted via the growth function as $\delta_{c}(z) = \delta_{c}(z=0) / D(z)$. This formulation encapsulates the redshift dependence of the process within the critical density contrast while incorporating mass dependence through the variance.

A more realistic collapse model can be used to analyze barrier crossing through a scale-dependent approach, inspired by the ellipsoidal collapse model (see Appendix \ref{app:I}). In the simple case of Markov trajectories related to a sharp-k space window function for smoothing the density contrast and spherical collapse model, the first-crossing statistics are given by

\begin{equation}
	f_{\rm{FU}}(S,\delta_c)=\frac{1}{\sqrt{2\pi}}\frac{\delta_c}{S^{3/2}}\exp\left[-\frac{\delta_c^2}{2S}\right]dS.
\end{equation}

The number density of structures can be obtained by the statistics of the first up-crossing
\begin{equation}
	n(M,t)dM=\frac{\bar{\rho}}{M}f_{\rm{FU}}(S,\delta_c)|\frac{dS}{dM}|dM.
\end{equation}
The evolution of large-scale structures from the linear to nonlinear regime gives rise to a clustered universe organized into a web-like configuration, referred to as the cosmic web. This structure consists of dark matter halos, filaments, sheets, and voids. Voids are underdense regions characterized by a density lower than the background density, in smoothed scales corresponding to void sizes. At late times, a significant fraction of the Universe's volume is occupied by voids. Remarkably, within the framework of the EST, an analytical estimation of their number density can be obtained. 

In the $\delta_L - S$ two-dimensional space, voids correspond to trajectories that first down-cross of the critical barrier for void formation, $\delta_v$, at a specific scale. Importantly, trajectories that up-cross the $\delta_c$ barrier prior to down-crossing are excluded. Hence, the voids embedded within halos structures, are not permanent structures and subsequently erased by halo collapse. The number of trajectories satisfying this condition was proposed in \cite{Sheth:2004hie}.

\be\label{eq:ffuvoid}
S\times{\cal{F}}(S,\delta_v,\delta_c) = \sum_{j=1}^{\infty} \frac{j^2\pi^2{\cal{D}}^2}{\delta_v^2/S}\frac{\sin(j\pi{\cal{D}})}{j\pi}\exp\left(-\frac{j^2\pi^2{\cal{D}}^2}{2{\delta_v}^2/S}\right),
\ee
where
\be\label{eq:Dvoid}
{\cal{D}}\equiv \frac{|\delta_v|}{(\delta_c + |\delta_v| )}.
\ee
Now, the number density of voids with radius $R$ and mass $m=\frac{4\pi}{3} R^3\bar{\rho}$ can be expressed in terms of the first down-crossing statistics as follows:
\be
\frac{m^2n_v(m)}{\bar{\rho}}=S\times{\cal{F}}(S,\delta_v,\delta_c) \frac{d\ln S}{d\ln m}.
\ee
To find the relation between the density contrast of each smoothing scale in the nonlinear regime, $\delta_{\text{NL}}$, and its linear counterpart utilized in the two-dimensional EST plane, $\delta_\text{L}(z)$, we employ the equation presented in \cite{Cooray:2002dia}.
\be
1+\delta_{\text{NL}} = \frac{m}{\bar{\rho}V_{\text{E}}} \approx \left(1-\frac{\delta_\text{L}(z)}{\delta_c}\right)^{-\delta_c}.\label{eq:NL-L}
\ee
For halos, the linear critical density value $\delta_{\rm{L}}(z) = \delta_c \simeq 1.69 $ corresponds to the $\delta_{\rm{NL}}\rightarrow \infty$. In the case of voids, the critical density contrast for void formation is $\delta_{\rm{L}}(z) = \delta_v\simeq -2.7$ which is obtained by setting $\delta_{\rm{NL}}=-0.8$. Due to mass conservation, the Eulerian radius of voids (the late-time nonlinear radius) is related to the Lagrangian radius (the early-time linear regime radius) via $R_E/R_L\simeq 1.7$.

One step further, EST can be employed to estimate the number of DM-halos within voids (halo-in-void) by considering multiple crossing scenarios. Trajectories that exhibit a first down-crossing at the scale $S_v(R)$ (void) from $\delta_v$, followed by a second up-crossing at the scale $S_h(M)$ (halo) at $\delta_c$, correspond to DM-halos embedded within voids. In the Markov version of the EST, this quantity can be computed straightforwardly by
\begin{eqnarray} \label{eq:ffuc}
	f_{\text{FU}}(S_h(M),\delta_c(z)|S_v(R),\delta_v(z))=\frac{1}{\sqrt{2\pi}}.\frac{\delta_c(z) - \delta_v (z)}{\left(S_h(M)-S_v(R)\right)^{3/2}} \\ \nonumber \times\exp\left({-\frac{(\delta_c(z) - \delta_v(z))^2 }{2(S_h(M)-S_v(R))}}\right).
\end{eqnarray}
It is important to note that the redshifts assigned to both void and halo formation are identical.

The above conditional probability calculation can be also applied to find the progenitor history of DM halos as 
\begin{eqnarray} \label{eq:ffuc}
f_{\text{FU}}(S_1,\delta_1|S_2,\delta_2)=\frac{1}{\sqrt{2\pi}}.\frac{\delta_c(z_1) - \delta_c (z_2)}{(S_1-S_2)^{3/2}} \\ \nonumber \times\exp\left({-\frac{(\delta_c(z_1) - \delta_c(z_2))^2 }{2(S_1-S_2)}}\right),
\end{eqnarray}
where $S_1$ and $S_2$ correspond to mass $M_1$ and $M_2$ respectively with condition $M_2>M_1$ and $z_2<z_1$. 
This probability calculation is also referred to as halo-in-halo or the mass assembly history of DM-halos in the literature.  
In the next section, we examine all the statistics introduced in this section in the presence of modified initial conditions.

\section{Results: Modified initial condition and non-linear structure formation}
\label{Sec:3}

In this section, we examine the impact of modified initial conditions on the non-linear formation and evolution of large-scale structures (LSS) at late time. We investigate the connection between these modifications and the statistical properties of Dark Matter (DM) halos and voids, as well as their evolutionary statistics within the Excursion Set Theory (EST) paradigm. These evolutionary statistics include halo-in-halo and halo-in-void number densities, which serve as a more sensitive and complementary probe to distinguish between modified models and the standard $\Lambda$CDM cosmological model.

In our previous work, we demonstrated that adding a simple Gaussian bump to the initial curvature power spectrum as a toy model could address the Too Big to Fail (TBTF) and missing satellite problems \citep{kameli:2020mod}. As discussed thoroughly in our earlier paper and the introduction of this manuscript, this initial power excess could arise from various inflationary scenarios. For pioneering works on inflationary models which altered the initial power spectrum, see \cite{Salopek:1988qh,starobinsky:1992spe,Randall:1995dj,Adams:1997de,Martin:2000xs,Hunt:2004vt}. Moreover, for more recent works involving the initial condition impacts on large-scale structure see \cite{hahn2011multi,agarwal2014cons} and for ideas involve initial non-Gaussianity and their impact of large-scale structure see \cite{slosar2008cons,desjacques2010pri,tseliakhovich2010nona}. 

In this paper, we do some complementary investigation on DM-halos statistics under these toy models. We study more realistic ellipsoidal collapse model for both Markov and non-Markov trajectories in EST context. We investigate different toy models to better understand the impact of initial condition. Furthermore, we will study the variation of Gaussian parameter and their effects on different large-scale structure statistics.

\begin{figure}
	\includegraphics[width=\columnwidth]{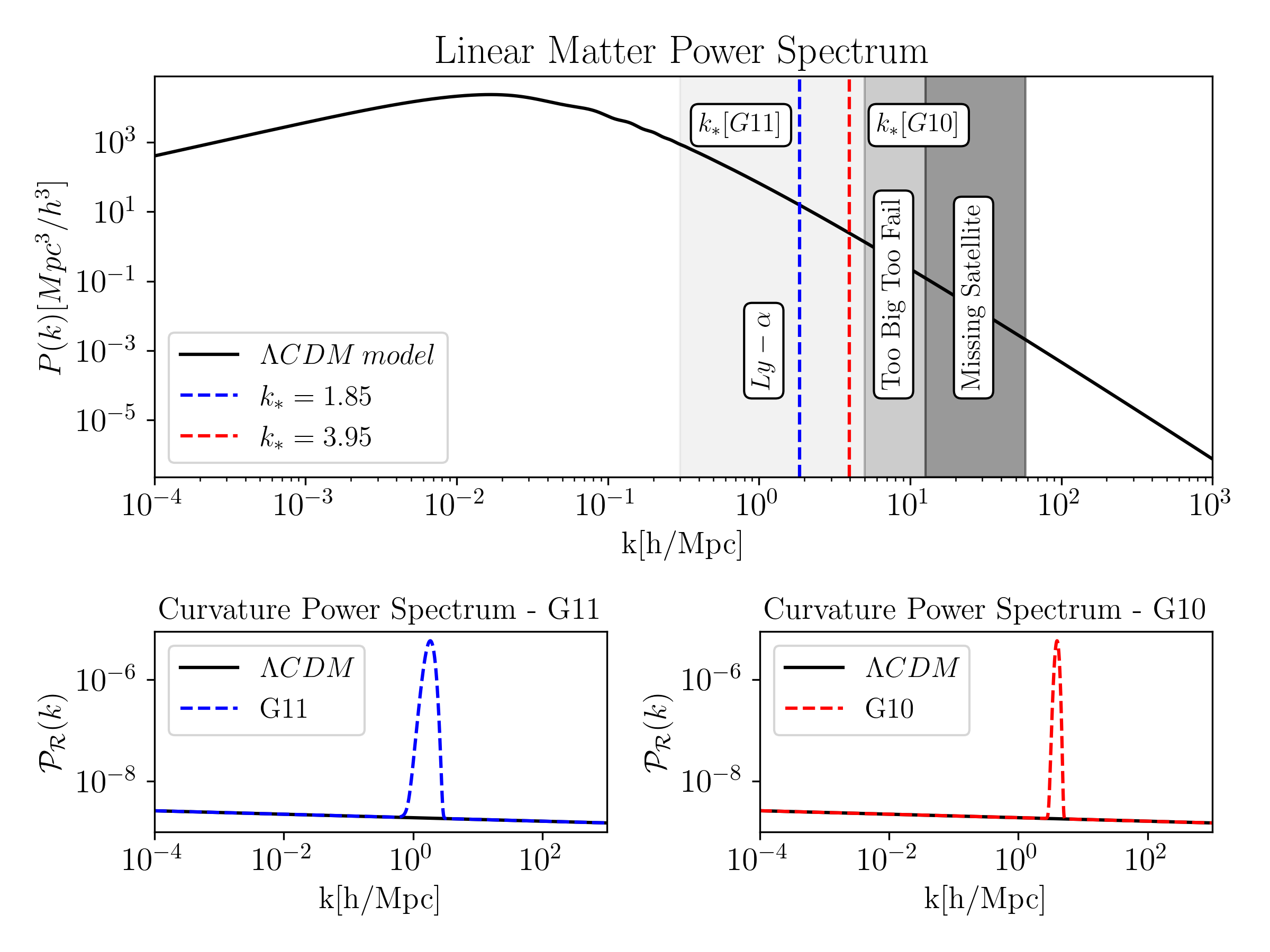}
	\caption{The linear matter power spectrum  is plotted versus the wavenumber (k) in the upper panel. The gray regions show the span of the wavenumbers of Ly-$\alpha$ observations, TBTF problem and missing satellite one. The blue/red-dashed vertical line show the $k_*[G11]$/$k_*[G10]$ values. In the bottom panels, we show the primordial curvature dimensionless power spectrum versus wavenumber for G11 (left panel) and G10 (Right panel). The modified Gaussian bump-like powers are shown with blue-dashed (G11) and red-dashed (G10) line.} \label{fig-ps}
\end{figure}

Additionally, we delve into the statistics of voids and halo-in-void. While most of the Universe's mass is confined in halos as virialized overdense regions, the majority of cosmic volume resides within underdense regions known as voids. The study of void properties and their statistical behavior is essential for a comprehensive understanding of the cosmic web and large-scale structure formation. This investigation extends to study the effects of initial conditions and inflationary features on void statistics. The cosmic web consist of both overdense halos and underdense voids, and initial modifications must affect both in a mutually consistent manner.

We hypothesize that changes in halo number densities will naturally alter void statistics, and such modifications must align with observational data for both halo and void. In our earlier study, we highlighted how initial power spectrum modifications could offer solutions to Too Big to Fail (TBTF) and the missing satellite problems. Here, we provide evidence suggesting that these modified models may simultaneously resolve the void phenomena problem, first introduced by \cite{peebles:2001void}.

For the initial curvature power spectrum modification, we propose a model wherein a simple Gaussian bump is added to the standard $\Lambda$CDM model. The modified curvature power spectrum is defined as:

\begin{equation}
{\cal{P}}^{\rm{MOD}}_{\cal{R}}(k)= \bar{{\cal{P}}}^{\Lambda \rm{CDM}}_{\cal{R}}(k)+{\cal{P}}^{\rm{bump}}_{\cal{R}}(k),
\end{equation}

where, ${\cal{P}}^{\rm{bump}}_{\cal{R}}(k)$ is a Gaussian function and parameterized as:    

\begin{equation}
{\cal{P}}^{\rm{bump}}_{\cal{R}}(k) = \frac{A_{b}}{\sqrt{2\pi} \sigma_b}\exp\left[-(k-k_*)^2/2\sigma^2_b\right],
\end{equation}

Here, $A_b$ denotes the amplitude of the Gaussian bump, $k_*$ represents the wavenumber at the center of the bump, and $\sigma_b$ corresponds to the variance of the bump.

\begin{figure*}
	\includegraphics[width=17.5cm]{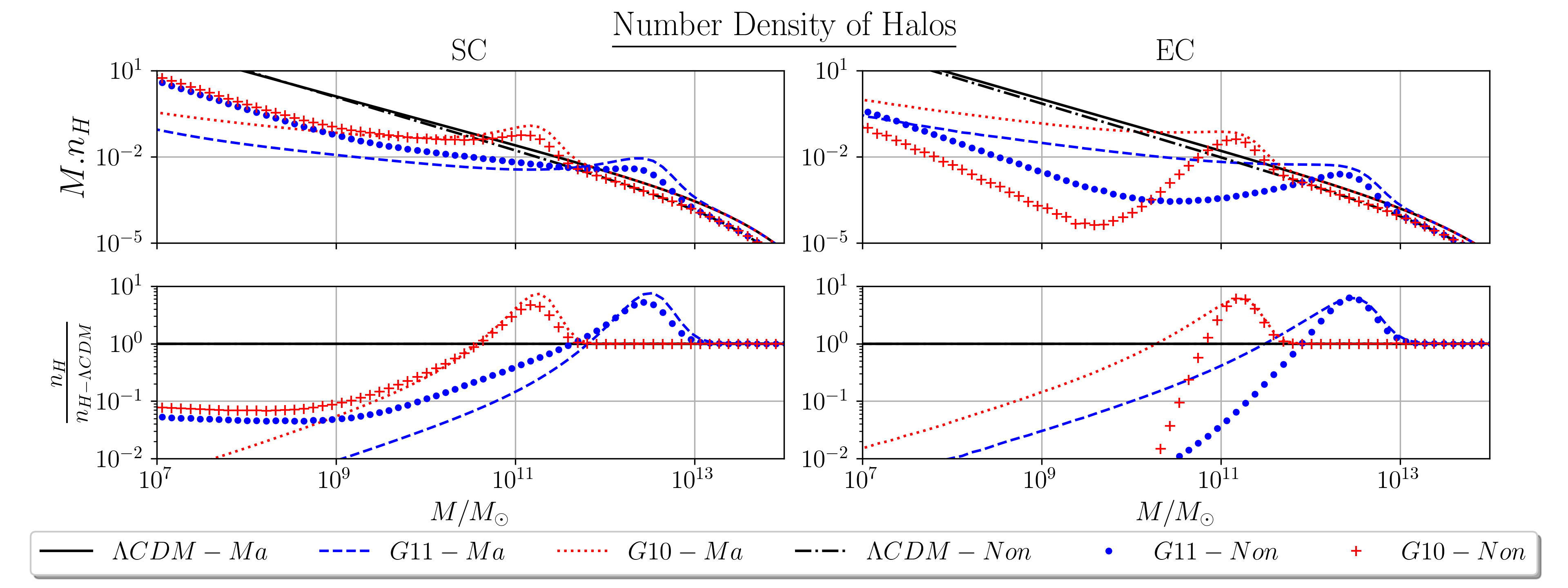}
	\caption{The number density of Dark Matter Halos (DM-Halos) versus DM-Halo mass at current time z=0, alongside their ratio to the standard $\Lambda$CDM model, is presented. Data are presented for both the standard $\Lambda$CDM model and the modified models G10 and G11. Results are shown for Markov (Ma) and non-Markov (Non) cases, as well as for Spherical Collapse (SC) and Ellipsoidal Collapse (EC). The upper panels depict the DM-Halo number density for different cosmological models, while the lower panels illustrate the ratio of different models to the standard $\Lambda$CDM model. The left panels correspond to Spherical Collapse (SC), and the right panels correspond to Ellipsoidal Collapse (EC). The standard model is indicated by black, the G11 model by blue, and the G10 model by red. } \label{fig:Halo}
\end{figure*}

\begin{figure*}
	\includegraphics[width=17.5cm]{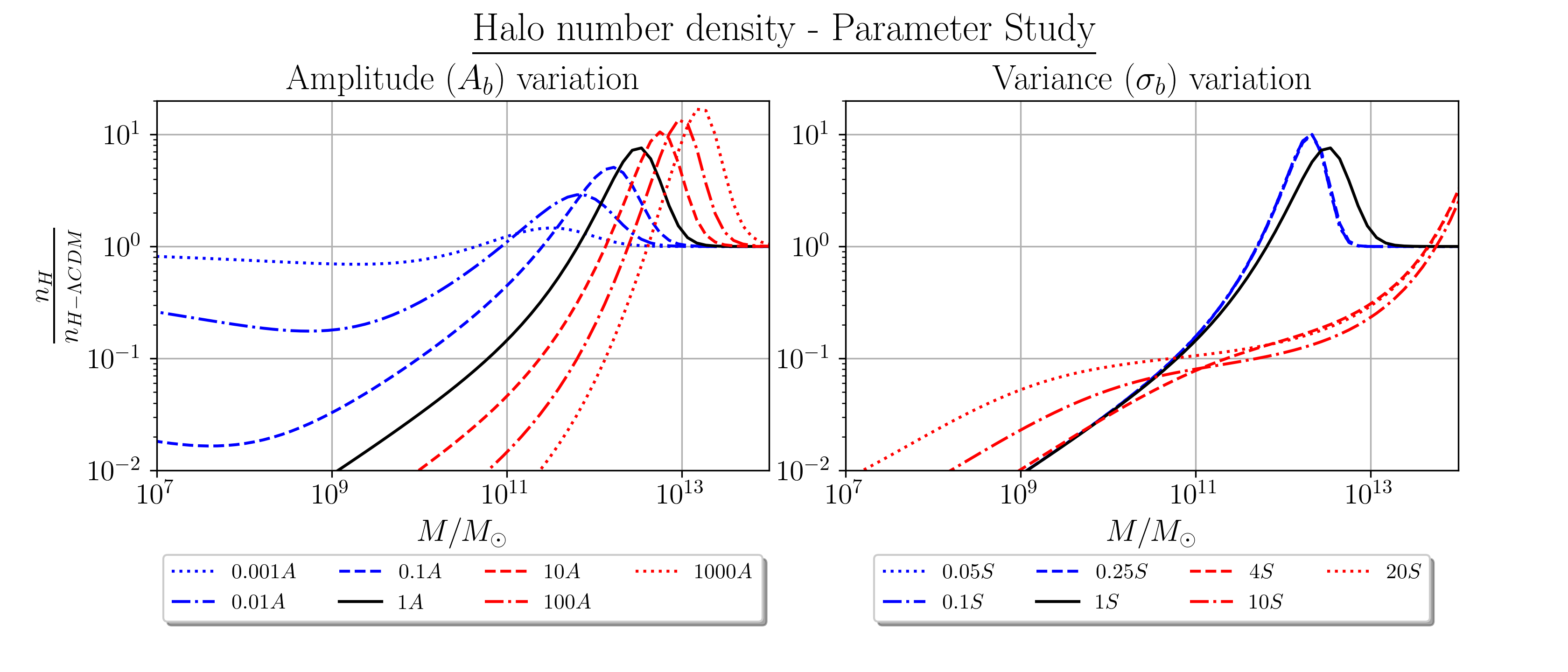}
	\caption{The ratio of the number density of DM-Halos to the standard $\Lambda$CDM model versus halo mass is shown for the G11 model with different Gaussian Bump parameters. All results correspond to the Markov-SC model, which has an exact analytical solution. In the left panel, results for various Amplitudes (A) of the Gaussian Bump modification are plotted, while in the right panel, results for different variances ($\sigma$) are plotted.  The black line represents the G11 model with its original parameters. Red curves indicate larger amplitude/variance, and blue curves represent smaller parameters.} \label{fig:HaloPS}
\end{figure*} 

The central wavenumber ($k_*$) of the Gaussian bump is the most critical parameter, as it corresponds to a specific mass scale that is expected to alter structure formation statistics. Notably, we will observe its cumulative impact on both larger and smaller mass scales in the following discussions. The matter power spectrum at large scales (small wavenumbers) is tightly constrained, while at small scales (wavenumbers $k \gtrsim 1 \, h/{\rm Mpc}$), it is only constrained by Ly-$\alpha$ observations. In our previous study, we demonstrated that the Ly-$\alpha$ constraints are consistent with the non-linear power spectrum derived from our modified model. For a detailed explanation, refer to Fig. 5 in \cite{kameli:2020mod}.

The linear matter power spectrum of the standard $\Lambda$CDM model is presented in the upper panel of Fig. \ref{fig-ps}. The shaded regions indicate the Ly-$\alpha$ constraints and the scales associated with the TBTF and missing satellite problems. Given these constraints at large-scales, we introduced two modified models, designated as G11 and G10. The blue and red dashed lines correspond to the $k_*$ scale of the G11 and G10 models, respectively. The G11 model corresponds to $k_* = 1.85$ h/Mpc, associated with a mass scale of $M_* = 10^{11} M_\odot$, while the G10 model corresponds to $k_* = 3.95$ h/Mpc, related to a mass scale of $M_* = 10^{10} M_\odot$. For both models, we apply an amplitude of $A_b = 2000$ and a variance of $\sigma_b = 0.25$. Considering the critical role of mass scales in cosmic structure formation, the only distinction between these two cosmological models lies in their $k_*$ values and, correspondingly, their associated mass scales. 
{{As highlighted in the introduction, our proposed method is also applicable to the issue of an excessive number of satellites. A larger $k_*$ than G10 results in a smaller-scale deficit and an excess within the mass range of massive dwarf galaxies. This presents a more natural solution compared to introducing a power deficit in the initial power spectrum, as such a feature can be realized in early universe models.}}

We will analyze the effects of bump amplitude and variance parameters on DM-halos and voids. The curvature power spectrum of the standard model and the modified models are illustrated in the bottom panels of Figure \ref{fig-ps}. The G11 model is depicted in blue-line in the right panel, while the G10 model is shown in red-line in the left panel. Throughout this manuscript, we use this color-coding convention (G11 in blue and G10 in red).

In the following subsections, we present and review the results of large-scale structure for the standard $\Lambda$CDM model, with Planck 2018 \citep{Aghanim:2018eyx} parameter (nearly scale-invariant curvature power spectrum) and the modified models G11 and G10, which we referred to as cosmological models. Additionally, we provide results for various technical models, which we referred to as EST-models. The statistical calculations for the EST-models are conducted for the following cases:
a) Markov trajectories (Ma) with spherical collapse (SC), b) Markov trajectories (Ma) with ellipsoidal collapse (EC), c) Non-Markov trajectories (Non) with spherical collapse (SC), and d) Non-Markov trajectories (Non) with ellipsoidal collapse (EC). Among these, the Ma-SC model is the only one, which have an analytical solution. Hence, this model is utilized for more detailed investigations and parameter studies, as computational cost for other models are significantly high. The theoretical foundations and concepts have discussed in section \ref{Sec:2}.

For the remaining EST-models (except for Ma-SC), we employ a computational approach. In our approach, we generate a large number of trajectories ($n_{traj} = 10^8$) for each EST-model, for both Markov and non-Markov trajectories. A computational technique is then applied to count the stochastic trajectories that first up-cross the collapse barrier. After the first up-crossing of the barrier, the counting process for each trajectory is terminated, thereby we avoided multiple halo counts per trajectory and eliminated the occurrence of halo-in-halo phenomena. This methodology is applied similarly to both halo and void computations. The halo collapse barrier is set at $\delta_c = 1.69$ for spherical collapse model, whereas the void barrier is set to $\delta_v = -2.71$, representing underdense void regions. For void calculations, trajectories that up-cross the collapse barrier before the void barrier are excluded. A comprehensive explanation of void calculations can be found in \cite{Sheth:2004hie,paranjape:2012hie}.

\begin{figure*}
	\includegraphics[width=17.5cm]{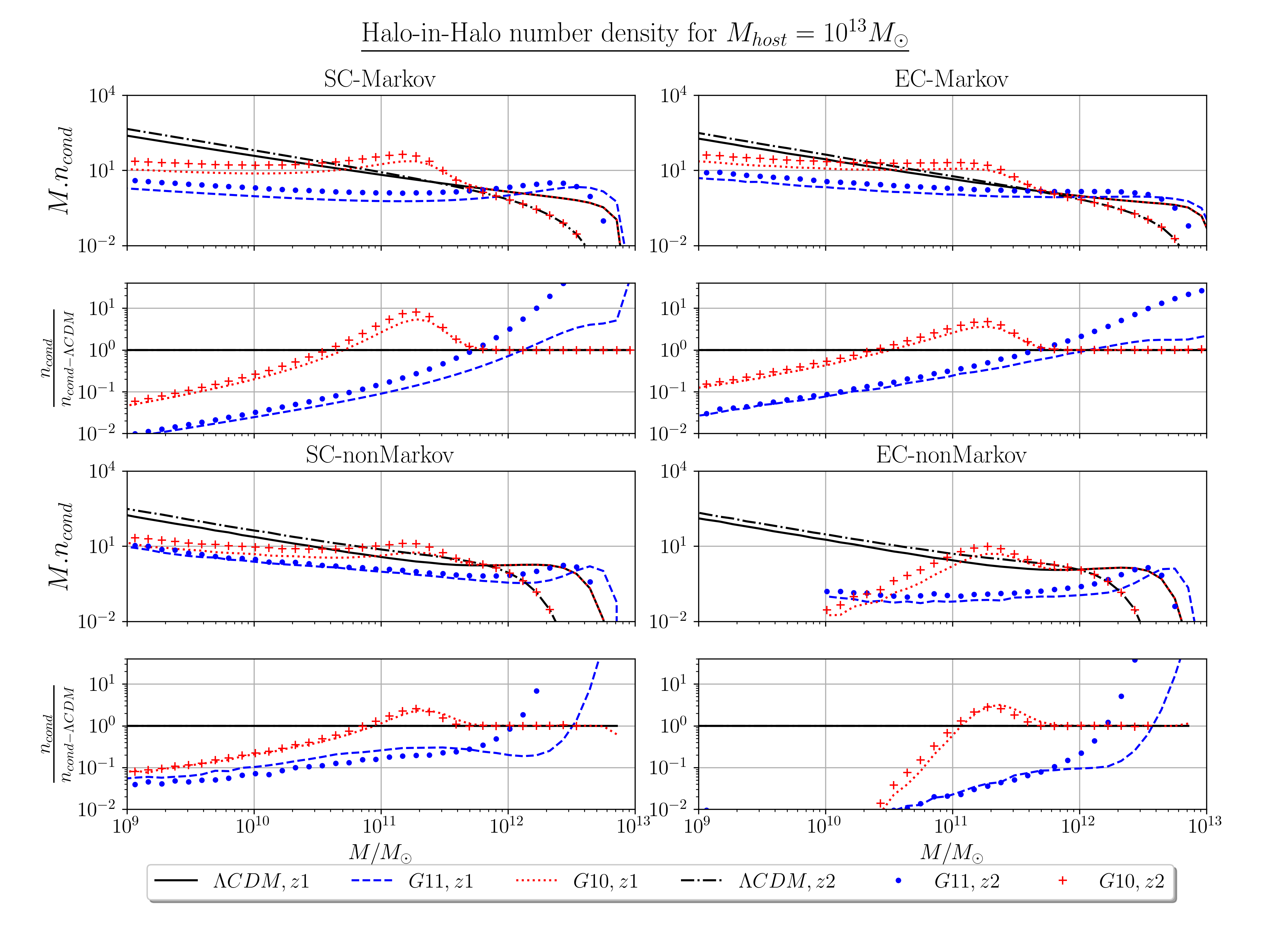}
	\caption{The conditional number density of Halo-in-Halo versus progenitor halo mass is depicted for both the standard $\Lambda$CDM model and the modified models G10 and G11, along with their ratio to the standard $\Lambda$CDM model. The conditional number density of progenitor halos at redshifts $z_1=1$ and $z_2=2$, which eventually form a host halo with $M=10^{13} M_\odot$ at $z_0=0$, is presented. Data are shown for both Markov/non-Markov cases and SC/EC collapse models. The upper two rows display the number density and its ratio for the Markov model, while the lower two rows correspond to the non-Markov models. The left panels represent Spherical Collapse (SC), and the right panels represent Ellipsoidal Collapse (EC). The standard model is indicated by black, the G11 model by blue, and the G10 model by red.  } \label{fig:HiH}
\end{figure*}

\begin{figure*}
	\includegraphics[width=17.5cm]{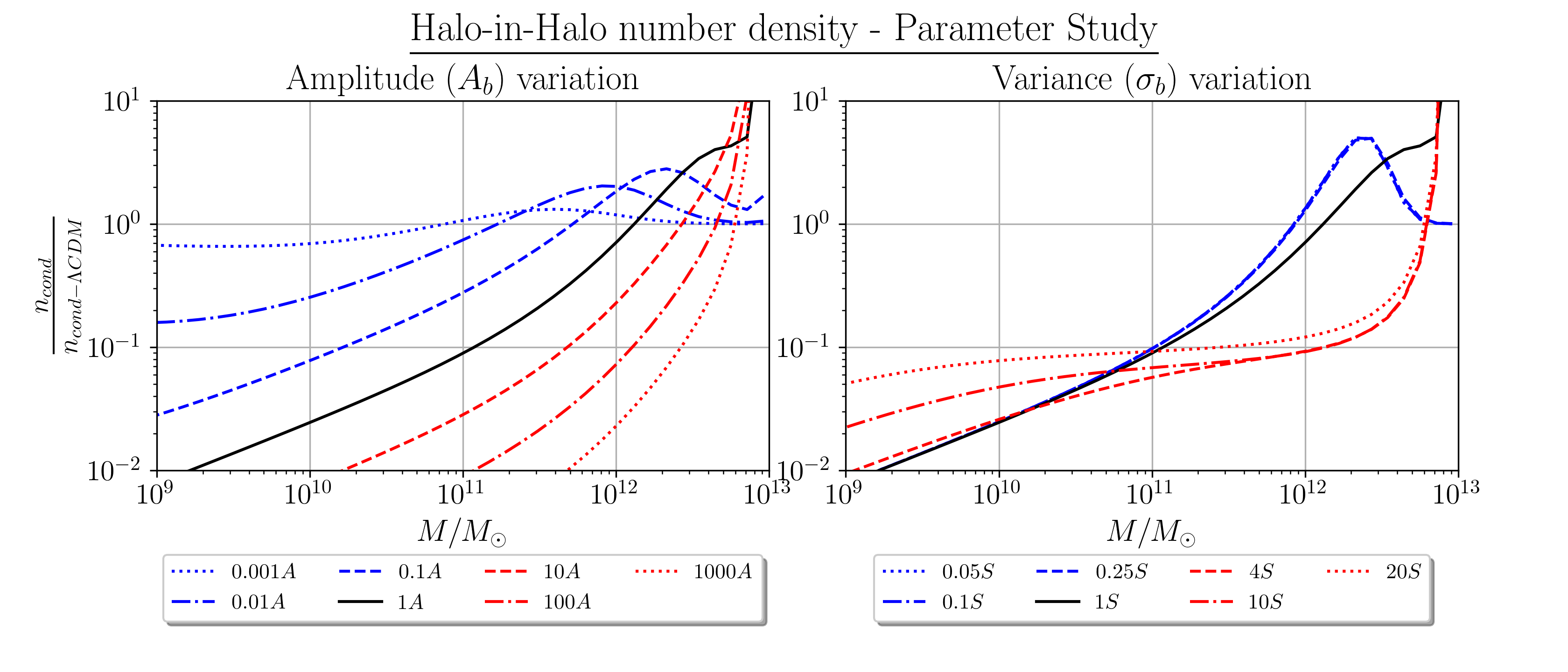}
	\caption{The ratio of the conditional number density of Halo-in-Halos to the standard $\Lambda$CDM model versus progenitor halo mass is shown for the G11 model with different Gaussian Bump parameters. The conditional number density of progenitor halos at redshifts $z_1=1$, which eventually form a host halo with $M=10^{13} M_\odot$ at $z_0=0$, is presented. All results correspond to the Markov-SC model, which has an exact analytical solution. In the left panel, results for various Amplitudes (A) of the Gaussian Bump modification are plotted, while in the right panel, results for different variances ($\sigma$) are plotted. The black line represents the G11 model with its original parameters. Red curves indicate larger amplitude/variance, and blue curves represent smaller parameters.} \label{fig:HiHPS}
\end{figure*}

The methodology for evolutionary statistics is somewhat more complicated. For halo-in-halo calculations, one must consider a specific time (current time, $z=0$ in our results) and a specific host halo mass (e.g., $M=10^{13} M_{\odot}$ in our results). These parameters correspond to a unique point in the trajectory's two-dimensional plane ($\delta_L-S$), see Fig. \ref{fig:traj}. The trajectories that first cross this specific point are identified and counted. Then, their second up-crossing from a redshift-dependent barrier ($\delta (z)$) is calculated at higher redshifts to determine the distribution of progenitor halos. Over time, these progenitor halos at earlier epochs (higher redshifts) gain mass through merger and accretion process to form larger host halos at the present time. Thus, conditional (evolutionary) statistics of DM-halos provide a comprehensive indicator of the mass assembly history of a final host halo with specified mass. 

The process for halo-in-void statistics follows a similar approach. However, for halo-in-void, the first crossing point relates to a void with a specified Eulerian radius at the current time ($z=0$). Subsequently, the mass distribution of embedded halos within the specified void is calculated. In this study, we examine voids with $R = 20$ and $30 \, {\rm Mpc}$. It should be reminded that Ma-SC is the only model for which an analytical solution exists.

In the following subsections, we present and discuss the results for four important statistics: 1- DM-halo number density, 2- Halo-in-Halo (conditional) number density or halo mass assembly history, 3- Void number density, and 4- Halo-in-Void number density. Results are provided for all four EST-models, consisting Markov and non-Markov trajectories and spherical and ellipsoidal collapses, as well as for three cosmological models: the standard model, G11, and G10.
      
\subsection{DM Halo Statistics}

In Fig. \ref{fig:Halo}, the DM-halo number density as a function of halo mass is presented for various cosmological and EST-models. The upper panels display the number density, while the lower panels show the ratio of the number density of each cosmological model to the standard $\Lambda$CDM model. In each ratio plot, a unique EST-model is applied consistently to both the modified and standard cosmological models. This approach is used throughout the other figures. For example, the blue-dashed plot in the right-lower panel represents the ratio of the G11 cosmological model to the standard $\Lambda$CDM model, under the non-Markov trajectories with an ellipsoidal collapse. By following this scheme, comparisons of different cosmological models remain almost independent of the chosen EST-model. To analyze the effects of different EST-models, comparisons are provided in Appendix \ref{app:II}. 

The key insight derived from the results in Fig. \ref{fig:Halo} is that an enhancement (or bump) in the initial conditions leads to a corresponding increase in the number density at mass scales near the modified model's related to $k_*$ value. The related mass scale for G11 (G10) is $M_{G11} = 10^{11} $ ($M_{G10} = 10^{10}) \  M_\odot$. However, we need more detailed analysis. The number density of halos in the modified models equals that of the standard model at a mass scale slightly larger than the corresponding $k_*$ mass scale. For the G11 (G10) model, this equality point is approximately $\sim 2.5 \times 10^{11}$ ($\sim 2.5 \times 10^{10}$) $M_\odot$. At mass scales exceeding this equality point, the modified models predict an increased number density of more massive DM-halos. The maximum difference in the number density occurs at mass scales approximately one order higher, $M\simeq 10^{12} M_\odot$ for the G11 model and $M\simeq10^{11} M_\odot$ for the G10 model, which we can see a lag effect here in comparison with initial bump. This amplified number density continue until mass scales two orders larger. After two order of mass the results for all cosmological models converge. As anticipated, these findings illustrate a Gaussian bump-shaped excess in the number density of massive halos. 

Additionally, a lag effect is observed in the results, a one-order magnitude discrepancy between the initial conditions and the late-time halo results. This lag arises from the cumulative enhancement of variance at specific scales. The variance, as defined in equation \ref{variance}, is computed by integrating the product of the matter power spectrum, which is enhanced by the initial curvature power, and a smoothing window function. This integration results in a gradual increase in variance at larger mass scales, thereby producing the lag effect, observed in the halo number density results.

The increased number density of massive halos comes at a tradeoff price. Given the Universe's constant total mass, this excess results in a significant reduction in the number density of low mass halos. Although only one order of magnitude more massive halos are observed at the peak within a relatively narrow mass window (spanning two orders of magnitude of mass), the logarithmic scale of DM halos reveals a reduction of one to two orders of magnitude in the number density of all smaller halos. This outcome provides a non-baryonic solution to the two critical small-scale challenges in the standard model: the Too Big to Fail (TBTF) problem and the missing satellite problem. These challenges refer to observations of too much fewer halos or satellite halos in N-body simulations and semi-analytical models compared to the observations. Observed small halos or dwarf galaxies are two orders of magnitude fewer than simulation predictions. Our proposed solution demonstrates that a modified initial condition can yield more massive halos while reducing sharply the number of low-mass halos relative to the standard model. 

The ratio of number densities for each modified model remain consistent for different EST-models. This consistency is particularly evident in the ratio plots (lower panels). Therefore, we conclude that the impact of different EST-models primarily serves as a correction to the analytical EST model (Markov-SC). These corrections are discussed in Appendix \ref{app:II}.

In Fig. \ref{fig:HaloPS}, variations in the parameters of Gaussian bump for the G11 model are examined. The number density ratio of the modified model to the standard model is depicted for different bump parameters. The left panel illustrates variations in the amplitude $A_b$, while the right panel presents variations in the variance $\sigma_b$. The black line represents the original G11 model. Results are provided for amplitudes $A_b = [0.001, 0.01, 0.1, 1, 10, 100, 1000] \times A_{b-G11}$ and variances $\sigma_b = [0.05, 0.1, 0.25, 1, 4, 10, 20] \times \sigma_{b-G11}$. Due to the high computational cost of other EST-model simulations, all parameter variation results are calculated by the analytical Markov-SC model. 

The amplitude variation exhibits a notable phenomenon. Despite a logarithmic variation spanning six orders of magnitude (from 0.001 to 1000), the maximum number density increases by only one order of magnitude. This indicates weak sensitivity of the results to the Gaussian bump amplitude. Furthermore, larger amplitudes shift the equality point (the mass scale the deviation of number density to standard model is started) toward higher halo masses, while the number density of low-mass halos decreases more dramatically. This weak sensitivity suggests that meaningful results may still arise in realistic scenarios with lower amplitude values. Hence, it would be intriguing to explore realistic features of inflation and other initial condition, such as initial non-Gaussian models.

The right panel of Fig. \ref{fig:HaloPS} illustrates the impact of variance variations of Gaussian bump. The number density ratio for higher variances (red lines) extends toward larger mass scales. While these larger mass scales are less relevant to our study, we conclude that toy models with broader variances at smaller scales ($k_*$) can generate comparable results within the mass ranges of our interest. Conversely, for narrower variances (blue lines), results exhibit uniformity and tend toward behavior resembling a Dirac Delta function as the variance decreases. This parameter study highlights the robustness of results across an extensive range of values and functional forms.


\subsection{Halo-in-Halo Statistics}

The halo-in-halo number density is depicted in Fig. \ref{fig:HiH}. This figure illustrates the number density of progenitor halos at redshifts $z_1=1$ and $z_2=2$, plotted versus progenitor halo mass. These progenitor halos gradually gain mass through accretion and merger processes, ultimately forming a host halo with a final mass of $M=10^{13} M_\odot$ at the present time ($z_0=0$). This calculation is alternatively referred to as the halo mass assembly history. Similar to the halo results, halo-in-halo statistics are computed for different EST-models. The ratio of the modified cosmological models to the standard $\Lambda$CDM model is presented in the lower panel for each EST-model.

\begin{figure*}
	\includegraphics[width=17.5cm]{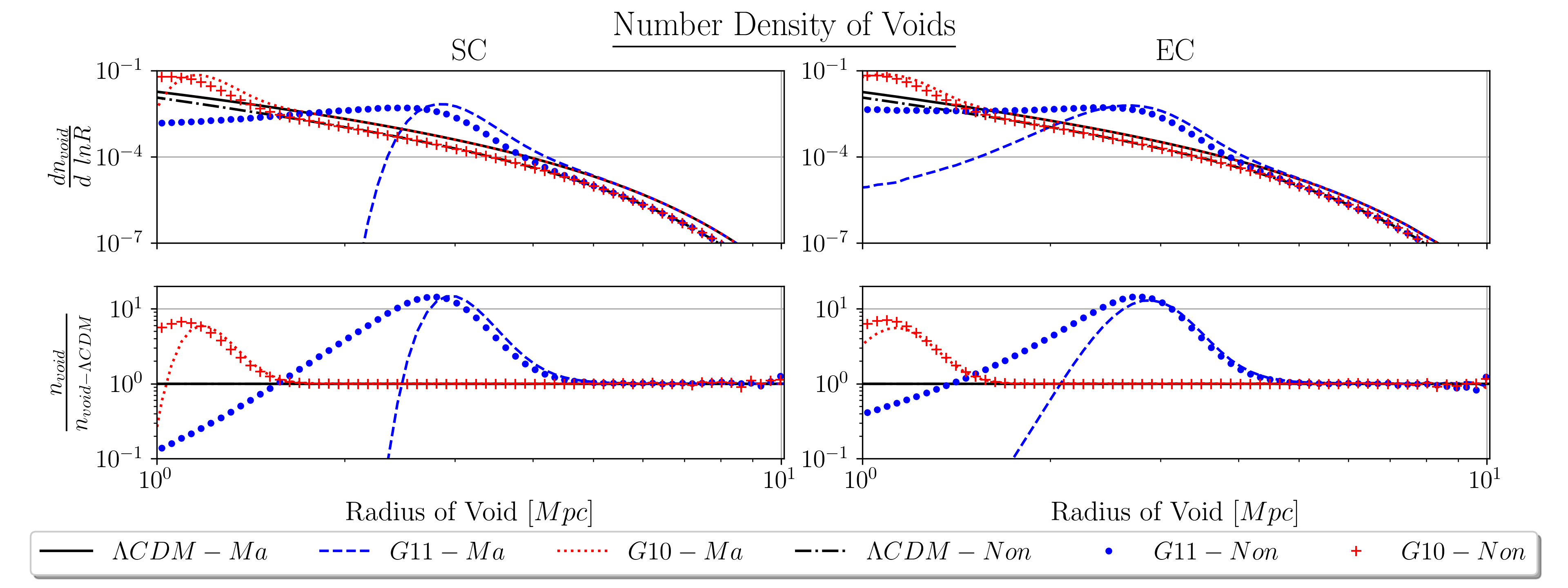}
	\caption{The number density of Voids versus Eulerian Radius at the current time $z=0$, along with their ratio to the standard $\Lambda$CDM model, is presented. Data are shown for both the standard $\Lambda$CDM model and the modified models G10 and G11. Results include Markov (Ma) and non-Markov (Non) cases, as well as Spherical Collapse (SC) and Ellipsoidal Collapse (EC). The upper panels show the Voids number density for different cosmological models, while the lower panels show the ratio to the standard $\Lambda$CDM model. The left panels represent Spherical Collapse (SC), and the right panels represent Ellipsoidal Collapse (EC). The standard model is shown in black, the G11 model in blue, and the G10 model in red. } \label{fig:Void}
\end{figure*}

\begin{figure*}
	\includegraphics[width=17.5cm]{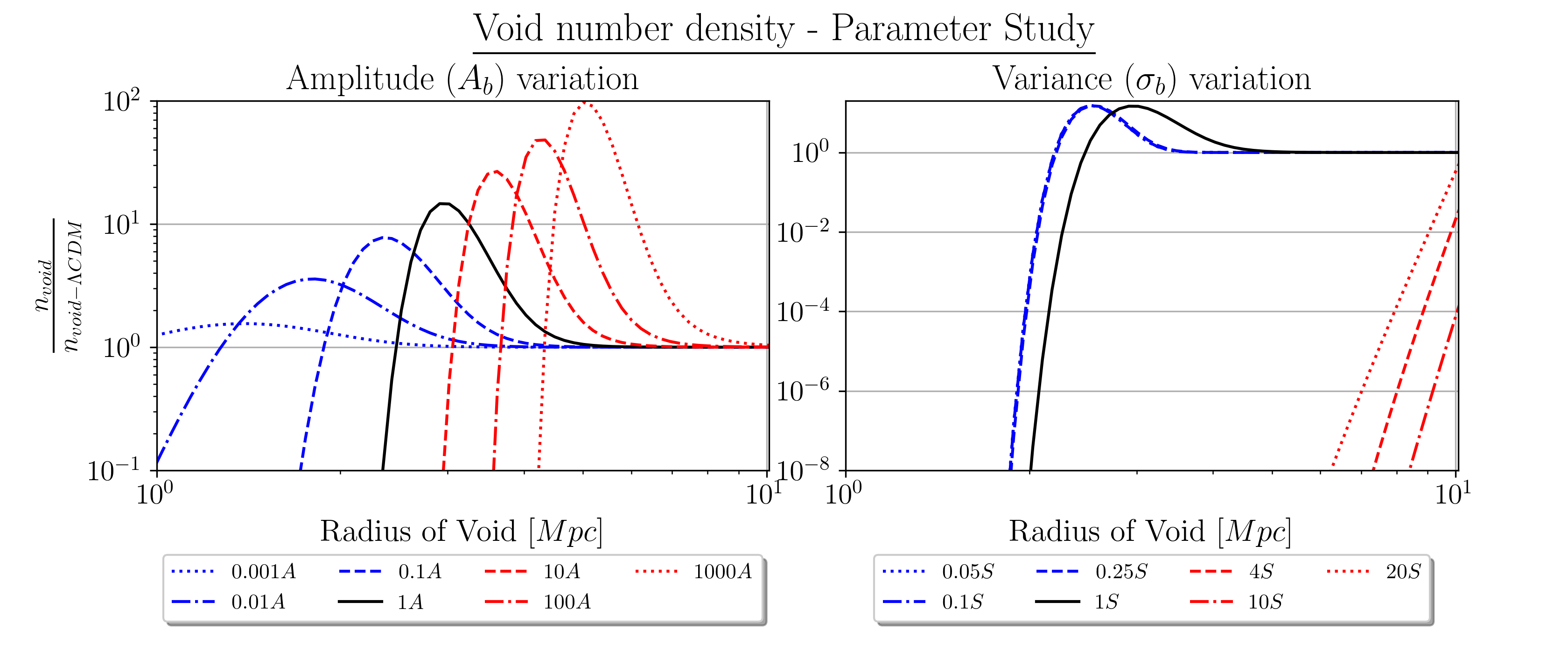}
	\caption{The ratio of the number density of Voids to the standard $\Lambda$CDM model versus Eulerian radius is shown for the G11 model with different Gaussian Bump parameters. All results correspond to the Markov-SC model, which has an exact analytical solution. In the left panel, results for various Amplitudes (A) of the Gaussian Bump modification are plotted, while in the right panel, results for different variances ($\sigma$) are plotted.  The black line represents the G11 model with its original parameters. Red curves indicate larger amplitude/variance, and blue curves represent smaller parameters.} \label{fig:VoidPS}
\end{figure*}

For the G10 model, an excess of progenitor halos is observed within the mass range $M_{\rm{excess}} \simeq [5 \times 10^{10}, 5 \times 10^{12}]$, with the maximum excess occurring at $M_{max} \simeq 3 \times 10^{11} M_\odot$. The G10 model exhibits a ratio of halo-in-halo number density that is almost similar to the halo number density results, though with slightly higher ratios. In contrast, the G11 model displays stronger ratio results relative to the halo number density. The anticipated peak at $M\simeq 10^{12} M_\odot$ is absent due to proximity to the final host halo mass. For masses greater than $M > 10^{12} M_\odot$, the ratio demonstrates a substantial increase, reflecting up to a 30 times higher likelihood of merger or mass accretion near the final host halo mass. This trend indicates an elevated probability of major mergers involving halos of nearly equal mass (within one order of mass ratio). Conversely, for small-mass progenitors, a significant reduction—up to one or two orders of magnitude—is observed in the progenitor halo number density.

In summary, the results for halo-in-halo number density resemble those of halo number density, albeit with higher ratios. Thus, halo-in-halo statistics provide an improved and complementary probe for distinguishing between cosmological models. The findings also underscore that modifications to initial conditions resulted in decreasing the numbers of small-mass progenitors and increasing the numbers of large-mass progenitors. Consequently, the merger rate of halos with nearly equal mass ratio increases significantly, while mergers with highly large mass ratios become less frequent. This supports our proposed resolution to the TBTF and missing satellite problems by reducing the number of small progenitors, or more precisely, satellites, which eventually merge into larger host halos.

The ratios for progenitor number density in non-Markov models (particularly G10) are marginally smaller than those observed in Markov models. The smoother trajectories associated with non-Markov processes lead to reduced rates of mergers. However, these EST-models do not fundamentally or conceptually alter the results. They serve only to provide corrections that enhance precision. A detailed discussion of these differences is provided in Appendix \ref{app:II}.

Fig. \ref{fig:HiHPS} presents the variations in the amplitude and variance of the Gaussian bump for the G11 model. The results for the G10 model are nearly identical. This figure displays the ratio of the modified model to the standard model as a function of progenitor mass for varying bump parameters. The parameter values for amplitude and variance are similar to those applied in Fig. \ref{fig:HaloPS}. Similar to previous findings, a weak sensitivity to amplitude variations is observed. For example, an enhancement of six orders of magnitude in amplitude leads to only one order amplification in the maximum ratio of the modified model. However, higher amplitudes (red lines) exhibit a greater tendency toward larger progenitor masses compared to lower amplitudes (blue lines), implying an increased likelihood of major mergers with higher amplitudes.

The variance variations exhibit behavior consistent with the halo results. Smaller values of variance (blue lines) yield similar plots that converge toward a Dirac Delta function as the variance approaches its zero limit. Conversely, larger variances (red lines) demonstrate a strong preference for higher mass values. These findings suggest that host halos preferentially form from more massive progenitors and possibly a greater portion of major merger counterparts.

\subsection{Void Statistics}

Fig. \ref{fig:Void} presents the results depicting the number density of voids as a function of their Eulerian radius (upper panel), and the ratio of void number density predicted by the modified models ratio to the standard $\Lambda$CDM model (lower panel). These calculations incorporate all the EST models discussed. In our approach for void calculations, trajectories crossing the collapse barrier at smaller variance should be excluded, thereby the spherical collapse (SC) and ellipsoidal collapse (EC) in void calculation refer to this exclusion. This scheme first was proposed by \cite{Parkavousi2023void}. 

The modified models significantly influence the void number density within the range $R_v\simeq [1,5]$ Mpc. There is an excess of voids observed around $R_v\simeq 3$ Mpc, accompanied by a sharp decline for smaller void sizes. Nevertheless, these scales are not relevant to observational studies, which predominantly focus on voids with sizes on the order of 10 Mpc or larger. Hence, the void number density results lack observational and statistical merit. The void number density variation, however, remains valuable for understanding the concept of impact of initial condition on the statistics on the cosmic web. Additional comparisons of EST-models within void statistics are presented in Appendix \ref{app:II}, serving purely computational and conceptual purposes.

In Fig. \ref{fig:VoidPS}, parameter study plots for different bump parameters are provided, using the same parameters as in the previous sections. Once again, the primary insight derived from these plots is a sharp decline in void number density for smaller voids, indicative of cutoff values at minimal radii. These findings, however, remain devoid of observational and statistical merit. As demonstrated in the left panel, a broader variance range (red lines) could potentially alter the statistical properties of the void radius distribution within the observationally relevant range and enhance the likelihood of detecting super-voids. This phenomena need further investigation, to test different initial condition impacts on emerging super-voids.

\begin{figure*}
	\includegraphics[width=17.5cm]{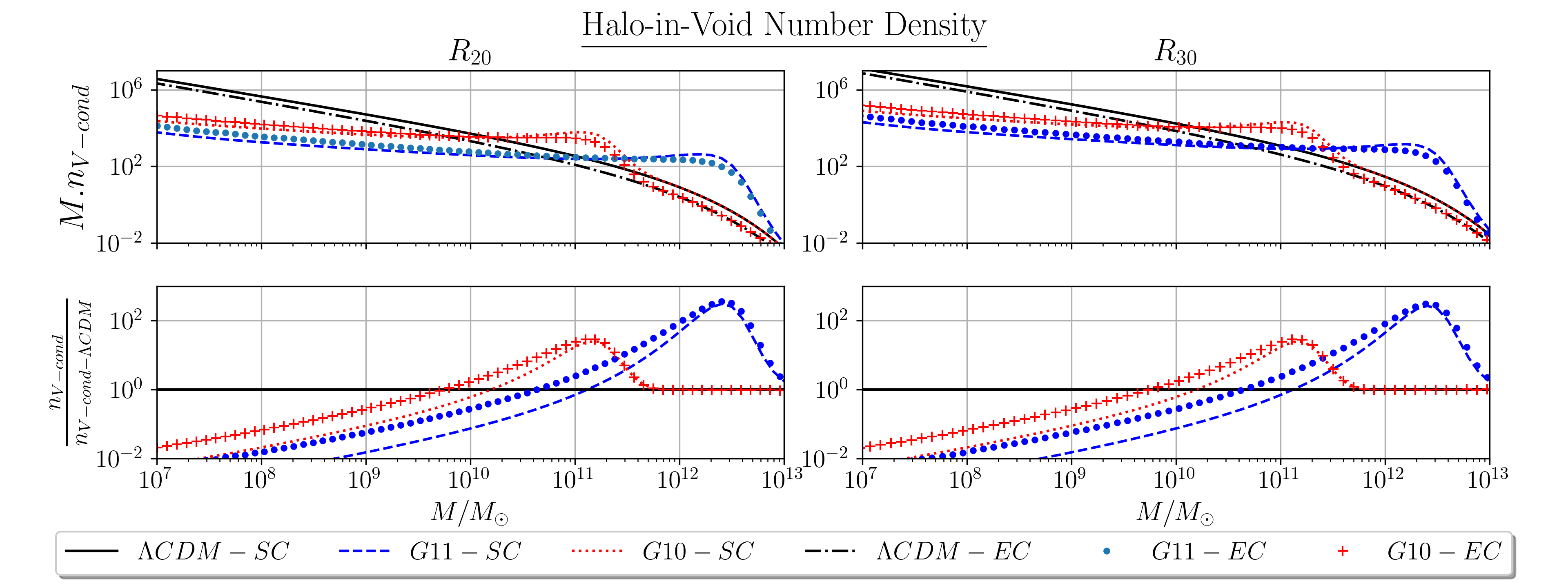}
	\caption{The conditional number density of Halo-in-Voids versus embedded halo mass is depicted for both the standard $\Lambda$CDM model and the modified models G10 and G11, along with their ratio to the standard $\Lambda$CDM model. The conditional number density of embedded halos within the Void with Eulerian radius $R_{20}=20 Mpc$ and $R_{30}=30 Mpc$ at current time $z_0=0$, is presented. Data are shown for Spherical Collapse (SC) and Ellipsoidal Collapse (EC) models. The upper panels display the number density and the lower panel show its ratio to standard $\Lambda$CDM model. The left panels represent Spherical Collapse (SC), and the right panels represent Ellipsoidal Collapse (EC). The standard model is indicated by black, the G11 model by blue, and the G10 model by red.  } \label{fig:HiV}
\end{figure*}

\begin{figure*}
	\includegraphics[width=17.5cm]{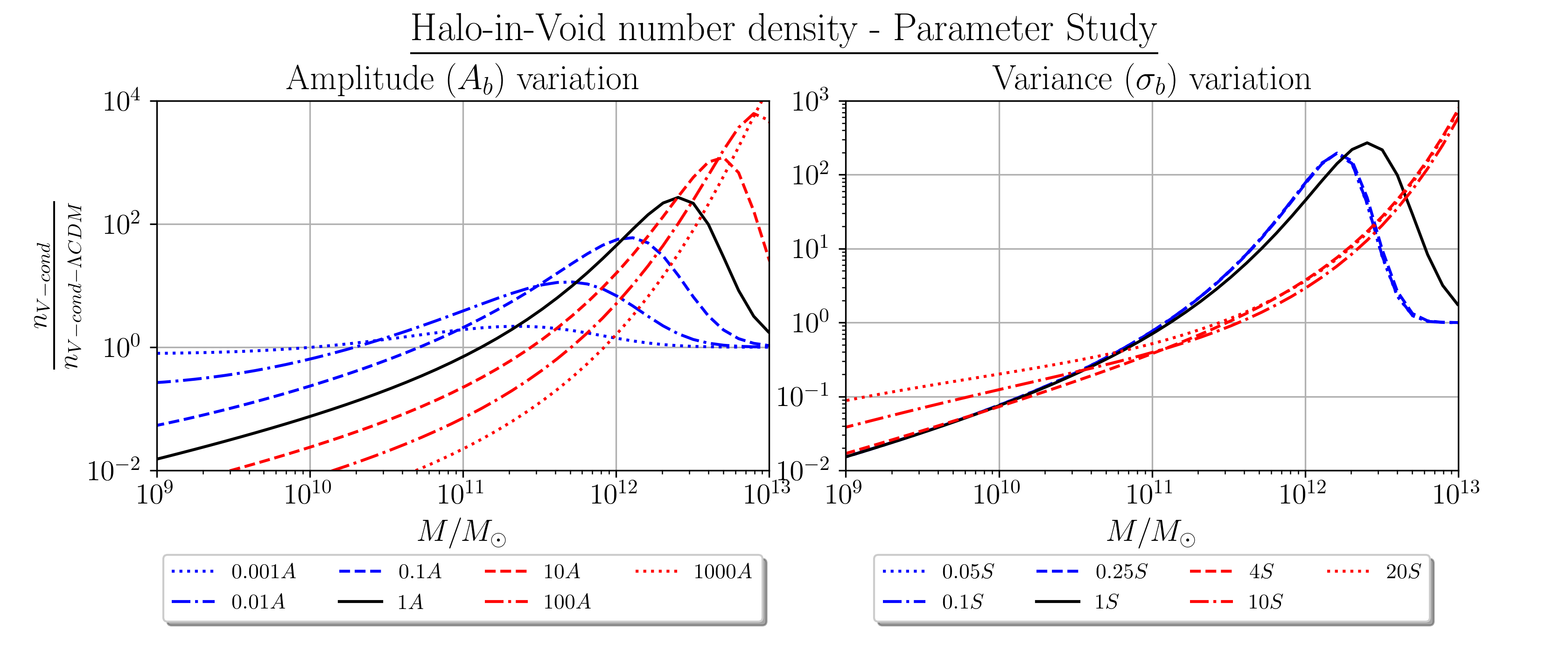}
	\caption{The ratio of the conditional number density of Halo-in-Voids to the standard $\Lambda$CDM model versus embedded halo mass is shown for the G11 model with different Gaussian Bump parameters. The conditional number density of halos within voids with Eulerian radius  $R_{30}=30 Mpc$ at redshifts $z_0=0$, is presented. All results correspond to the Markov-SC model, which has an exact analytical solution. In the left panel, results for various Amplitudes (A) of the Gaussian Bump modification are plotted, while in the right panel, results for different variances ($\sigma$) are plotted. The black line represents the G11 model with its original parameters. Red curves indicate larger amplitude/variance, and blue curves represent smaller parameters. } \label{fig:HiVPS}
\end{figure*}

\subsection{Halo-in-Void Statistics}

The final and most informative dataset is the halo-in-void statistics. Fig. \ref{fig:HiV} illustrates the number density of halos embedded in voids with specific radius at the present epoch ($z=0$). Results are shown for two distinct void radii, $R_{20}=20 \, \mathrm{Mpc}$ (left panels) and $R_{30}=30 \, \mathrm{Mpc}$ (right panels). The halo-in-void number density is calculated for the standard model, as well as for modified models, using Markov trajectories within the frameworks of SC and EC Collapse. A brief comparison of SC and EC models is provided in Appendix \ref{app:II}.

It should be noted that, due to technical limitations and prohibitively high computational costs, results for non-Markov trajectories could not be obtained. Our computational methodology for conditional statistics involves identifying trajectories that initially cross the void negative barrier at a specific variance scale (corresponding to the void radius), followed by the second crossing of the collapse barrier (SC or EC). The second crossing determines the halo mass distribution embedded within voids of specific radius. This process is challenged by the fact that only a small fraction of trajectories cross the void barrier, necessitating an enormous number of generated trajectories to achieve meaningful statistics for their second crossing. Indeed, a significant proportion of trajectories are excluded during the first void barrier crossing. For Markov trajectories, characterized by uncorrelated Brownian motion, the first crossing point for a specific void scale can be straightforwardly utilized. This involves a simple coordinate transformation in the $\delta_L-S$ plane of origin to the specific point of first crossing. Conversely, non-Markov trajectories are fully correlated, requiring the correlation of all scales in trajectories. Consequently, the aforementioned coordinate transformation cannot resolve this issue for non-Markov cases.

Fig. \ref{fig:HiV} reveals a pattern similar to halo and halo-in-halo statistics, characterized by a bell-shaped excess in the number density of halos embedded in specific voids for both G10 and G11 models. The excess for the G11 (G10) model corresponds to an embedded halo mass range of $M_{G11} \simeq [10^{11}, 10^{13}] \, (M_{G10} \simeq [10^{10}, 5 \times 10^{11}] )\, M_\odot$, with maximum ratios observed at $M_{G11} \simeq 3 \times 10^{12} \, M_\odot$ and $M_{G10} \simeq 2 \times 10^{11} \, M_\odot$. While these findings are similar to the halo number density results, the maximum ratio here is approximately two orders of magnitude greater, establishing halo-in-void statistics as a more effective and sensitive discriminator among other statistics. It suggest, each void appears to function as a distinct universe, within which embedded halos exhibit behavior analogous to DM-halos, albeit with a heightened sensitivity to modifications in the initial conditions.

From an observational view, these results suggest a reduced number of low-mass embedded halos within voids, coupled with an elevated ratio of massive embedded halos. This interpretation aligns with the "void phenomenon" proposed by \cite{peebles:2001void}, which proposed that voids are not entirely empty. The faint galaxies are expected to inhabit in voids. Small DM-halos within these voids constitute a substantial mass fraction, which could showed by N-body simulations and our semi-analytical EST approach. The proposed solution addresses this by enhancing the number density of massive halos while significantly reducing the density of small halos and accordingly, faint galaxies. Observationally, this implies prioritizing searches for massive halos within voids, to searching for the faint galaxies, which decrease largely in modified models. Considering the logarithmic mass scale, these massive halos exclude a substantial fraction of small halos. Although, the number density of massive progenitors are enhanced by two orders of magnitudes compared to the standard model, yet massive halos remain rare events even within modified models as demonstrated in the figure upper panels. 

In Fig. \ref{fig:HiVPS}, the sensitivity of the halo-in-void statistics to amplitude and variance is explored. The analysis retains the same parameters and models as prior sections. Here, halo-in-void statistics exhibit remarkable sensitivity, with a four-order magnitude increase in number density corresponding to six order of amplitude variations. By comparison, other probes, such as halo and halo-in-halo number densities, show only a one-order increase across six orders of amplitude variation. This underscores halo-in-void statistics as the most effective indicator for validating or ruling out modified models. As a key finding, we propose utilizing halo-in-void statistics and comparing them with observational data to investigate initial models and inflationary features. The ratio of hal-in-void number density for voids with radus 30 Mpc is 3-5 times greater that the void with radius of 20 Mpc.


\section{Conclusion and Future remarks}
\label{Sec:5}

This manuscript provides a complementary investigation to our previous study \citep{kameli:2020mod} into the impact of initial conditions on the nonlinear formation and evolution of large-scale structures as well as its impact on void statistics. Specifically, we assess the effects of introducing a Gaussian bump, as a toy model, into the initial curvature power spectrum on DM-halo and void statistics, as well as evolutionary statistics, halo-in-halo and halo-in void. We adopt Excursion Set Theory (EST), incorporating extensions to account for non-Markov trajectories and ellipsoidal collapse as more realistic models. Only the traditional EST formulation, Markov-SC admits an analytical solution. Therefore, we employed a computational approach for the extended EST variants. In practice, this involved counting the first up-crossing events for simple statistics (halo and void number densities) and the second crossing events for evolutionary probes (halo-in-halo and halo-in-void). As discussed in Appendix \ref{app:II}, these extended EST models serve as corrections to the analytical solution, and our results indicate that the ratios of the modified models to the standard case remain almost similar under the different EST implementations.

This study investigates the influence of modified initial conditions on large-scale structure formation, demonstrating their potential to address key small-scale challenges, including the Too Big to Fail problem, the missing satellite problem, and the void phenomenon. 
{{Although many baryonic solutions have been proposed in the literature to address the small-scale challenges of the standard model, we aim to present and emphasize a new perspective: a non-baryonic solution rooted in early universe physics.}} \\We present the results for modification impacts on the number densities of DM-halo, halo-in-halo, void, and halo-in-void observable. Our findings reveal that these statistics serve as effective probes for distinguishing between the modified models and the standard cosmological $\Lambda$CDM model. The standard $\Lambda$CDM model refers to standard initial condition, with nearly Gaussian, isotropic, adiabatic and nearly scale-invariant perturbation based on Planck 2018 \citep{Aghanim:2018eyx}. The evolutionary statistics, particularly halo-in-void, emerge as especially sensitive indicators. In other words, we propose using large-scale structure statistics as a diagnostic tool to validate or rule out alternative cosmological models and initial conditions. \\
{{In this context, developing a comprehensive observational strategy offers a promising pathway for advancing our understanding of large-scale structures in the Universe. The rapid and extensive growth of galaxy surveys—such as the Dark Energy Survey (DES), and upcoming projects like the Vera C. Rubin Observatory —opens up new opportunities to analyze the statistical properties of these structures in various cosmic environments.
		One particularly valuable approach involves constructing detailed catalogues of galaxies residing in cosmic voids. By studying the distribution, properties, and dynamics of galaxies within these voids, we can derive crucial statistical measures. These measures serve as sensitive tests for the predictions made by different cosmological models.
		Given that our proposed research focuses heavily on the behavior and characteristics of halos within voids, it presents a unique opportunity to directly test and validate our theoretical models against observational data from these large-scale surveys. The ability to compare our predictions with real-world measurements in void environments will not only strengthen the credibility of our models but also contribute to a more deep understanding of the role that these underdense regions play in the evolution of the cosmic web. }} \\
The toy modified models G11 and G10, implemented by adding a Gaussian bumps into the curvature power spectrum, address both the Too Big to Fail (TBTF) and the missing satellite problems by enhancing the number density of massive halos while reducing that of small halos by two to three orders of magnitude. Moreover, these modifications increase the probability for a host halo to accrete mass and merge at scales comparable to its own mass. Thereby, we have more major mergers and fewer minor ones for high-ratio mergers. This issue imply a reduced abundance of satellite halos and galaxies. Furthermore, the modified initial conditions shed light on the void phenomenon by revealing a consistent mutual effect between overdense regions (halos) and underdense regions (voids). Indeed, one of our most significant findings is that these modified initial condition models may offer a unified solution to both the small-scale halo challenges (TBTF and missing satellites) and the void phenomenon. Any modification that alters halo statistics is inevitably will change the void properties and their statistics.
{{A comment on the methodology is essential.  In particular, adopting a differentiable approach that enables gradient-based reconstruction of the primordial power spectrum from late-time observational data could be a promising direction to extend the toy model. Such methods could leverage recent advances in inverse modelling and machine learning, allowing us to directly optimize the primordial parameters by minimizing discrepancies between predicted and observed late-time measurements. However, it should be noted that the complex and non-linear nature of the EST model brings this line of thought out of the scope of this work and is a promising avenue for further study.
Regarding the Gaussian-bump parameterization, which we suggest, it is worth mentioning that it is a useful initial model that can serve as a flexible, parametric starting point within this broader framework. Its ability to capture localized features in the power spectrum makes it a convenient basis for initial testing, and it can be extended or modified as needed to incorporate more complex features.
	Moving forward, one can plan to explore how these parameterizations can be integrated into a differentiable pipeline, possibly employing techniques such as automatic differentiation and Bayesian inference, to enable robust and efficient reconstruction of the primordial spectrum directly from observational data. This would provide a more rigorous and quantitative pathway to connect early-universe physics with late-time measurements.}} \\
An additional key result of our study is that evolutionary statistics, including halo-in-halo and halo-in-void number densities, constitute highly sensitive probes. For example, while the overall halo number density increases by only one order of magnitude when the bump amplitude is raised by six orders, the halo-in-void number density experiences an increase of four orders under the same conditions. This distinctive sensitivity to variations in the Gaussian bump parameters positions these evolutionary observables, halo-in-void statistics in particular, as powerful discriminators for different cosmological models.

Our detailed key findings for each observable are summarized as follows: 

\textbf{DM Halos:} 
	The modified models (G10 and G11) exhibit a Gaussian-shaped enhancement in the number density of massive halos near their characteristic mass scales (related to $k_*$), approximately $10^{12}\, M_\odot$ for G11 and $10^{11}\, M_\odot$ for G10, while simultaneously reducing the abundance of smaller halos. This effect, which arises from cumulative variance enhancements, also reveals a slight lag between the modification scale and the formation mass scale of massive halos.
	
\textbf{Halo-in-Halo:} 
	Analysis of halo-in-halo statistics shows that modified initial conditions boost the number of progenitor halos in specific mass ranges, thereby increasing the likelihood of major mergers near the host halo mass. For instance, the G11 model exhibits up to a 30-times enhancement in progenitor densities at masses just above $10^{12}\, M_\odot$.
	
\textbf{Void and Halo-in-Void Statistics:} 
	Although the void number density shows enhancements only at small scales (around 3 Mpc), which is not of observational interest, the halo-in-void statistics are particularly striking. The modified models lead to an increase in the number density of halos embedded within voids, with maximum ratios up to two orders of magnitude higher than those in the standard model. Parameter studies further indicate that halo-in-void probe are exceptionally sensitive to variations in the Gaussian bump, making them a robust tool for differentiating between cosmological models. 

As suggestion for future investigations, it will be of considerable interest to test these modifications using more robust N-body simulations, as such studies could yield more detailed and informative datasets. While our semi-analytical approach offers a fast and reasonably accurate tool for modeling large-scale structure formation and evolution, more realistic methodologies are necessary for acquiring finer details. Moreover, although the halo-in-halo number density encompasses both merger and accretion processes, the merger rate, particularly that of major mergers, remains of greater observational relevance and importance. Quantifying the merger rate of DM-halos across different mass scales is a primary objective for our future works, especially since such major mergers may be observed via pair-galaxies observations.

Finally, we suggest extending this approach to more realistic initial conditions and alternate cosmological models. Future studies may involve the exploration of different inflationary models and their characteristic features, the role of initial non-Gaussianity, the influence of primordial black holes, and other primordial phenomena on large-scale structure. Additionally, we propose more investigation on the impact of initial condition on the formation of super-voids and the cold-spot phenomena. Our developed EST engine is well-suited to test various scenarios and straightforwardly assess their impact on late-time observations.

\section*{Data availability}
This study does not involve the use or production of original data.

\section*{Acknowledgments}
We thank the anonymous referee, whose insightful comments and detailed suggestions elevated the manuscript to a new level.\\
SB is partially supported by the Abdus Salam International Center for Theoretical Physics (ICTP) under the regular associateship scheme.
Moreover, SB are partially supported by the Sharif University of Technology Office of Vice President for Research under Grant No. G4010204. The authors utilized AI-assisted tool (Microsoft Copilot) to enhance language clarity and grammatical precision in this manuscript. \\



\begin{thebibliography}{99}


\bibitem[\protect\citeauthoryear{Adame et al.}{2025}]{2025JCAP...02..021A} Adame A.~G., Aguilar J., Ahlen S., Alam S., Alexander D.~M., Alvarez M., Alves O., et al., 2025, JCAP, {\bf{2025}}, 021. 


\bibitem[\protect\citeauthoryear{Adams et al.}{1997}]{Adams:1997de}
Adams J.~A.~, Ross G.~G.~ and Sarkar S.~, 1997,
Nucl.\ Phys.\ B {\bf 503}, 405.

\bibitem[\protect\citeauthoryear{DESI Collaboration et al.}{2016}]{2016arXiv161100036D} DESI Collaboration, Aghamousa A., Aguilar J., Ahlen S., Alam S., Allen L.~E., Allende Prieto C., et al., 2016, arXiv, arXiv:1611.00036. 



\bibitem[\protect\citeauthoryear{Planck Collaboration et al.}{2020}]{Aghanim:2018eyx} Planck Collaboration, Aghanim N., Akrami Y., Ashdown M., Aumont J., Baccigalupi C., Ballardini M., et al., 2020, A\&A, {\bf{641}}, A6. 


\bibitem[\protect\citeauthoryear{Agarwal et al.}{2014}]{agarwal2014cons}
Agarwal, Nishant and Ho, Shirley and Shandera, Sarah, 2018, , J. Cosmol. Astropart. Phys.,  {\bf 2014}, no. 2, 038.



\bibitem[\protect\citeauthoryear{Alam et al.}{2017}]{Alam:2016hwk}
Alam S. {\it et al.} [BOSS Collaboration], 2017,  MNRAS,  {\bf 470}, no. 3, 2617.







\bibitem[\protect\citeauthoryear{Baghram et al.}{2014}]{Baghram:2014nha}
Baghram  S., Abolhasani A.~A., Firouzjahi H. and Namjoo M.~H., 2014.   J. Cosmol. Astropart. Phys. {\bf 1412}, 036.

\bibitem[\protect\citeauthoryear{Baghram et al.}{2019}]{2019PhRvE..99f2101B} Baghram S., Nikakhtar F., Tabar M.~R.~R., Rahvar S., Sheth R.~K., Lehnertz K., Sahimi M., 2019, PhRvE, {\bf{99}}, 062101. 









\bibitem[\protect\citeauthoryear{Blum et al.}{2022}]{2022arXiv220307220B} Blum B., Digel S.~W., Drlica-Wagner A., Habib S., Heitmann K., Ishak M., Jha S.~W., et al., 2022, arXiv:2203.07220. 





\bibitem[\protect\citeauthoryear{Bond et al.}{1991}]{Bond:1990iw}
Bond   J.~R., Cole S., Efstathiou G. and Kaiser N., 1991,   ApJ,  {\bf 379}, 440 .


\bibitem[\protect\citeauthoryear{Boylan-Kolchin et al.}{2011}]{BoylanKolchin:2011de}
Boylan-Kolchin  M., Bullock J.~S. and Kaplinghat M., 2011,  MNRAS, {\bf 415}, L40.


\bibitem[\protect\citeauthoryear{Bullock et al.}{2017}]{Bullock:2017xww}
Bullock  J.~S. and Boylan-Kolchin M., 2017,  Ann.\ Rev.\ Astron.\ Astrophys.\  {\bf 55}, 343.








\bibitem[\protect\citeauthoryear{Chen et al.}{2016}]{2016JCAP...11..014C} Chen X., Dvorkin C., Huang Z., Namjoo M.~H., Verde L., 2016,  J. Cosmol. Astropart. Phys., {\bf{2016}}, 014. 


\bibitem[\protect\citeauthoryear{Cooray \& Sheth}{2002}]{Cooray:2002dia}
Cooray   A. and Sheth R.~K., 2002,  Phys.\ Rept.\  {\bf 372}, 1.





\bibitem[\protect\citeauthoryear{D'Amico et al}{2011}]{DAmico2011void}
D’Amico, G., Musso, M., Norena, J., \& Paranjape, A. 2010,  Phys. Rev. D, {\bf 83}, no. 2, 023521.

\bibitem[\protect\citeauthoryear{Di Valentino et al.}{2021}]{2021CQGra..38o3001D} Di Valentino E., Mena O., Pan S., Visinelli L., Yang W., Melchiorri A., Mota D.~F., et al., 2021, CQGra, {\bf{38}}, 153001.


\bibitem[\protect\citeauthoryear{Desjacques \& Seljak}{2010}]{desjacques2010pri}
Desjacques, Vincent and Seljak, Uro, 2010,  Advances in Astronomy, {\bf 2010}, no. 1, 808640.


\bibitem[\protect\citeauthoryear{Elgaroy et al.}{2002}]{Elgaroy:2001wu}
Elgaroy O., Gramann M. and Lahav O., 2002,   MNRAS  {\bf 333}, 93.


\bibitem[\protect\citeauthoryear{Eisenstein \& Hu}{1998}]{Eisenstein:1997ik}
Eisenstein  D.~J. and Hu W., 1998, ApJ  {\bf 496}, 605.






\bibitem[\protect\citeauthoryear{Fard \& Baghram}{2018}]{Fard:2017oex}
Fard M.~A. and Baghram S., 2018,   , J. Cosmol. Astropart. Phys., {\bf 1801}, 051 .


\bibitem[\protect\citeauthoryear{Fritz et al.}{2019}]{2019A&A...623A.129F} Fritz T.~K., Carrera R., Battaglia G., Taibi S., 2019, A\&A, 623, A129. doi:10.1051/0004-6361/201833458




\bibitem[\protect\citeauthoryear{Garrison-Kimmel et al.}{2014}]{Garrison-Kimmel:2014kia}
Garrison-Kimmel S., Horiuchi S., Abazajian K.~N., Bullock J.~S. and Kaplinghat M., 2014, MNRAS  {\bf 444}, no. 1, 961.


\bibitem[\protect\citeauthoryear{Hahn \& Abel}{2011}]{hahn2011multi}
Hahn, Oliver and Abel, Tom, 2011, MNRAS  {\bf 415}, no. 3, 2101.


\bibitem[\protect\citeauthoryear{Hassani et al.}{2016}]{Hassani:2015zat}
Hassani  F., Baghram S. and Firouzjahi H.~, 2016,   , J. Cosmol. Astropart. Phys. {\bf 1605}, 044.

\bibitem[\protect\citeauthoryear{Homma et al.}{2024}]{2024PASJ...76..733H} Homma D., Chiba M., Komiyama Y., Tanaka M., Okamoto S., Tanaka M., Ishigaki M.~N., et al., 2024, PASJ, 76, 733. 



\bibitem[\protect\citeauthoryear{Hunt et al.}{2004}]{Hunt:2004vt}
Hunt P.~ and Sarkar S.~,2004,
, Phys. Rev. D ,{\bf 70}, 103518.




\bibitem[\protect\citeauthoryear{Jennings et al}{2013}]{Jennings2013void}
Jennings, Elise, Yin Li, and Wayne Hu., 2013
MNRAS {\bf 434}, no. 3, 2167. 

\bibitem[\protect\citeauthoryear{Kameli \& Baghram}{2020}]{kameli:2020mod}
Kameli Hamed, and Baghram Shant, 2020,
MNRAS {\bf 494}, no. 4, 4907.

\bibitem[\protect\citeauthoryear{Kameli \& Baghram}{2022}]{kameli:2022Hten}
Kameli Hamed, and Baghram Shant, 2022
MNRAS  {\bf 511}, no. 2, 1601.

\bibitem[\protect\citeauthoryear{Kim, Peter, \& Hargis}{2018}]{2018PhRvL.121u1302K} Kim S.~Y., Peter A.~H.~G., Hargis J.~R., 2018, PhRvL, 121, 211302. doi:10.1103/PhysRevLett.121.211302


\bibitem[\protect\citeauthoryear{Klypin et al.}{1999}]{Klypin:1999uc}
Klypin A.~A., Kravtsov A.~V., Valenzuela O. and Prada F., 1999,   ApJ.,  {\bf 522}, 82.


\bibitem[\protect\citeauthoryear{Laureijs et al.}{2011}]{EUCLID:2011zbd}
Laureijs R. \textit{et al.} [EUCLID] , 2011.   [arXiv:1110.3193 [astro-ph.CO]].




\bibitem[\protect\citeauthoryear{Leo et al.}{2018}]{Leo:2017wxg}
Leo  M., Baugh C.~M., Li B. and Pascoli S.~, 2018.   , J. Cosmol. Astropart. Phys., {\bf 1808}, 001.


\bibitem[\protect\citeauthoryear{Ma et al}{2011}]{Ma2011non}
Ma, C. P., Maggiore, M., Riotto, A., \& Zhang, J., 2011, MNRAS  {\bf 411.4}, 2644.



\bibitem[\protect\citeauthoryear{Maggiore \& Riotto}{2010}]{maggiore2010non}
Maggiore, Michele, and Antonio Riotto, 2010, ApJ. {\bf 711.2}, 907.




\bibitem[\protect\citeauthoryear{Martin et al.}{2001}]{Martin:2000xs}
Martin J.~ and Brandenberger R.~H., 2001, Phys.\ Rev.\ D {\bf 63}, 123501.




\bibitem[\protect\citeauthoryear{Moore et al.}{1999}]{Moore:1999nt}
Moore B., Ghigna S., Governato F., Lake  G.~, Quinn T.~R., Stadel J.~ and Tozzi P.~, 1999, ApJ.  {\bf 524}, L19.




\bibitem[\protect\citeauthoryear{  Musso \& Sheth}{2014a}]{Musso:2014non1}
Musso M. and Sheth R.~K., 2014,  MNRAS {\bf 439}, no. 3, 3051.


\bibitem[\protect\citeauthoryear{  Musso \& Sheth}{2014b}]{Musso:2014non2}
Musso M. and Sheth R.~K., 2014, MNRAS {\bf 443}, no. 2, 1601.


\bibitem[\protect\citeauthoryear{Nakama et al.}{2017}]{Nakama:2017ohe}
Nakama T., Chluba J.~ and Kamionkowski M., 2017, Phys.\ Rev.\ D {\bf 95}, no. 12, 121302.


\bibitem[\protect\citeauthoryear{Namjoo et al.}{2014}]{Namjoo:2014nra}
Namjoo   M.~H., Abolhasani A.~A., Baghram S. and Firouzjahi H., 2014, , J. Cosmol. Astropart. Phys. {\bf 1408}, 002.




\bibitem[\protect\citeauthoryear{Nikakhtar \& Baghram}{2017}]{Nikakhtar:2016bju}
Nikakhtar F. and Baghram S., 2017, Phys.\ Rev.\ D {\bf 96}, no. 4, 043524.

\bibitem[\protect\citeauthoryear{Nikakhtar et al.}{2018}]{Nikakhtar:2018qqg}
Nikakhtar F., Ayromlou M., Baghram S., Rahvar S., Rahimi Tabar M.~R. and Sheth R.~K.~, 2018,  MNRAS {\bf 478}, no. 4, 5296.





\bibitem[\protect\citeauthoryear{Paranjape et al.}{2012}]{paranjape:2012hie}
Paranjape, Aseem, Tsz Yan Lam, and Ravi K. Sheth., 2012,  MNRAS {\bf 420.2}, 1648.

\bibitem[\protect\citeauthoryear{Paranjape, Lam, \& Sheth}{2012}]{2012MNRAS.420.1429P} Paranjape A., Lam T.~Y., Sheth R.~K., 2012, MNRAS, {\bf{420}}, 1429. 




\bibitem[\protect\citeauthoryear{Parkavousi et al.}{2023}]{Parkavousi2023void}
Parkavousi, Laya, Hamed Kameli, and Shant Baghram., 2023,  MNRAS  {\bf 526.1}, 1495.


\bibitem[\protect\citeauthoryear{Peebles}{2001}]{peebles:2001void}
Peebles, P. J. E., 2001, ApJ.,  {\bf 557}, 495.




\bibitem[\protect\citeauthoryear{ Press \& Schechter}{1974}]{Press:1973iz}
Press W.~H. and Schechter P.~,  1974,  ApJ.  {\bf 187}, 425.







\bibitem[\protect\citeauthoryear{Randall et al.}{1996}]{Randall:1995dj}
Randall L.~, Soljacic M.~ and Guth A.~H.~, 1996,
Nucl.\ Phys.\ B {\bf 472}, 377.


\bibitem[\protect\citeauthoryear{Riess et al.}{2019}]{2019ApJ87685R} Riess A.~G., Casertano S., Yuan W., Macri L.~M., Scolnic D., 2019, ApJ, {\bf{876}}, 85. 

\bibitem[\protect\citeauthoryear{Sales, Wetzel, \& Fattahi}{2022}]{2022NatAs...6..897S} Sales L.~V., Wetzel A., Fattahi A., 2022, NatAs, 6, 897. 



\bibitem[\protect\citeauthoryear{Salopek et al.}{1989}]{Salopek:1988qh}
Salopek D.~, Bond J.~ and Bardeen J.~M.~, 1989,
Phys.\ Rev.\ D {\bf40}, 1753.

\bibitem[\protect\citeauthoryear{Santos-Santos et al.}{2025}]{2025MNRAS.540.1107S} Santos-Santos I.~M.~E., Frenk C.~S., Navarro J.~F., Cole S., Helly J., 2025, MNRAS, 540, 1107. 

\bibitem[\protect\citeauthoryear{Sawala et al.}{2016}]{2016MNRAS.457.1931S} Sawala T., Frenk C.~S., Fattahi A., Navarro J.~F., Bower R.~G., Crain R.~A., Dalla Vecchia C., et al., 2016, MNRAS, 457, 1931.









\bibitem[\protect\citeauthoryear{Sheth, Mo, \& Tormen}{2001}]{2001MNRAS.323....1S} Sheth R.~K., Mo H.~J., Tormen G., 2001, MNRAS, 323, 1. 

\bibitem[\protect\citeauthoryear{Sheth \& Tormen}{2002}]{Sheth:2001dp}
Sheth  R.~K. and Tormen G., 2002, MNRAS  {\bf 329}, 61.


\bibitem[\protect\citeauthoryear{Sheth \& Weygaert}{2004}]{Sheth:2004hie}
Sheth, Ravi K., and Rien Van De Weygaert., 2004, MNRAS {\bf 350}, 517.

\bibitem[\protect\citeauthoryear{Slosar et al.}{2008}]{slosar2008cons}
Slosar, An{\v{z}}e and Hirata, Christopher and Seljak, Uro{\v{s}} and Ho, Shirley and Padmanabhan, Nikhil, 2008,  J. Cosmol. Astropart. Phys.,  {\bf 2008}, no. 08, 031.





\bibitem[\protect\citeauthoryear{Starobinsky et al.}{1992}]{starobinsky:1992spe}
Starobinsky A.A. , 1992,
JETP.\ lett.{\bf55}, 489.


\bibitem[\protect\citeauthoryear{Tavasoli}{2021}]{2021ApJ...916L..24T} Tavasoli S., 2021, ApJL, {\bf{916}}, L24. 


\bibitem[\protect\citeauthoryear{Tian et al.}{2025}]{2025A&A...696L..19T} Tian H., Liu C., Xue X.-X., Fan D., Luo C., Nie J., Yang M., et al., 2025, A\&A, 696, L19. doi:10.1051/0004-6361/202453610


\bibitem[\protect\citeauthoryear{Tinker \& Conroy}{2009}]{Tinker:2009void}
Tinker, Jeremy L., and Charlie Conroy., 2009,  ApJ.,  {\bf 691.1}, 633.




\bibitem[\protect\citeauthoryear{Tseliakhovich et al.}{2010}]{tseliakhovich2010nona}
Tseliakhovich, Dmitriy and Hirata, Christopher and Slosar, An{\v{z}}e, 2010, Phys. Rev. D, {\bf 82}, no. 4, 043531.








\bibitem[\protect\citeauthoryear{Weinberg et al.}{2015}]{2015PNAS..11212249W} Weinberg D.~H., Bullock J.~S., Governato F., Kuzio de Naray R., Peter A.~H.~G., 2015, PNAS, {\bf{112}}, 12249. 


\bibitem[\protect\citeauthoryear{Weltman et al.}{2020}]{2020PASA...37....2W} Weltman A., Bull P., Camera S., Kelley K., Padmanabhan H., Pritchard J., Raccanelli A., et al., 2020, PASA, {\bf{37}}, e002. 





\bibitem[\protect\citeauthoryear{Zentner}{2007}]{Zentner:2006vw}
Zentner A.~R., 2007, Int.\ J.\ Mod.\ Phys.\ D {\bf 16}, 763.





\end{thebibliography}


\begin{twocolumn}
	
	\appendix
	\section{EST extensions} \label{app:I} 
We employ non-Markov trajectories and the ellipsoidal collapse model as more realistic extension of the EST. These extensions serve as corrective modifications to the conventional EST framework, which use Markov trajectories and spherical collapse.

\subsection{Ellipsoidal Collapse}

Within the framework of the EST, ellipsoidal collapse corresponds to a variance-dependent collapse barrier. The curved ellipsoidal collapse barrier at two distinct redshifts is illustrated in Fig. \ref{fig:traj}. In this manuscript, we adopt a computational approach to count the first up-crossing of the collapse barrier by trajectories. Thus, the sole distinction between the spherical collapse (SC) and ellipsoidal collapse (EC) models lies in the differing heights of their respective barriers.

The ratio of the critical density barrier in the ellipsoidal collapse model, $\delta_\text{EC}$, to that of spherical collapse model is as below \cite{2001MNRAS.323....1S,Sheth:2001dp}: 

\begin{equation} \label{eq:ellip-barrier}
	\frac{\delta_\text{ec}}{\delta_\text{sc}} = \sqrt{\bar{a}} [1+\beta (\bar{a}\nu)^{-\alpha}],
\end{equation}
where $\bar{a}\approx 0.7$, $\alpha \approx 0.615$ and $\beta \approx 0.485$. 

The curved ellipsoidal collapse barrier is applied across all collapse models to evaluate various number densities, including halo, halo-in-halo, and halo-in-void number density. In void calculations, the ellipsoidal collapse (EC) model accounts for the exclusion of trajectories that up-cross the collapse barrier before reaching the void’s negative barrier \cite{Parkavousi2023void}. The ellipsoidal collapse model is applicable for both Markov and non-Markov trajectories.  
\subsection{Non-Markov extension}

In the standard EST framework, we use a specific window function, known as the k-space sharp filter. This window function serves to smooth density perturbations within a spherical region of radius \( R \) centered at position \( x \).
\be
\delta_L(x,R)=\int \frac{d^3 k}{(2\pi)^3} \delta(k) W(k,R)e^{-ik.x}.
\ee

The k-space sharp filter $W(k,R)$ induces an uncorrelated random Markov walk. This same window function can also be used for variance calculations, as outlined in equation \ref{variance}. As discussed in detail in the theoretical section, this choice of window function in the Markov framework admits an analytical solution.

More realistic smoothing functions, such as the Gaussian filter and the real-space top-hat window function, introduce correlated steps in the density contrast random walk. In a Markov process, each step $\delta_L$ depends solely on the previous step, meaning the trajectory retains no memory of past steps. In contrast, alternative window functions give rise to correlated trajectories, where all prior steps influence subsequent evolution.

To construct these correlated trajectories using the top-hat or Gaussian window functions, we employ the methodology introduced by \cite{Nikakhtar:2018qqg} and extended to realistic cosmological models by \cite{kameli:2020mod}. This approach utilizes a numerically exact algebraic technique, Cholesky decomposition, to generate correlated non-Markov trajectories \citep{2019PhRvE..99f2101B}. The Cholesky method provides a robust numerical framework ensuring that the generated trajectories exhibit the correct ensemble properties. The statistical characteristics of random walks depend on the correlation structure across all individual steps. To systematically generate these correlations across different scales, we introduce the correlation matrix $C_{ij}$, which encodes the density contrast heights of the trajectory at any two distinct scales.
\be
\langle \delta_i\delta_j\rangle\equiv C_{ij} = \int \frac{dk}{k}\frac{k^3P_m(k)}{2\pi^2}\tilde{W}(kR_i)\tilde{W}(kR_j),
\ee
where $P_m(k)$ is the linear matter power spectrum and $\tilde{W}(kR_i) [\tilde{W}(kR_j)]$ is the window function in the smoothing scale $R_i  [R_j]$ respectively. The diagonal indices ($C_{ii}=S_i$) are equal to variance (S) at different scales. The height of the walk for each step $n$ defined as
\be
\delta_{n}=\langle \delta_n | \delta_{n-1}, ... ,\delta_1 \rangle + \sigma_{n|n-1, ..., 1}\xi_n ,
\ee
\begin{figure}
	\includegraphics[width=\columnwidth]{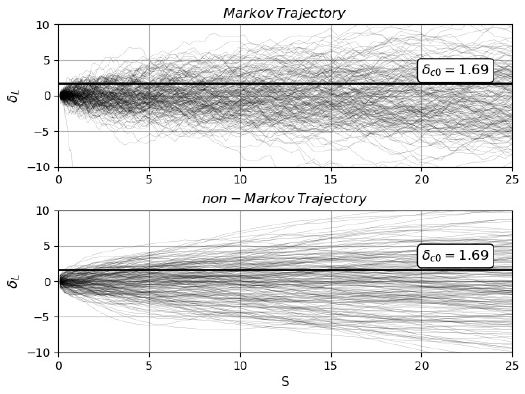}
	\caption {Comparison between Markov (upper panel) and non-Markov (lower panel) trajectories \protect\citep{kameli:2020mod}. The non-Markov trajectories exhibit  smoother shape due to their inherent correlation across multiple scales.  
	}\label{fig:nonmarkov}
\end{figure}

The first term represents the dependence of the height at the $n$th step on all preceding scales, while the second term show its correlation with the smoothed variance. The variable $\xi_n$ is a Gaussian random number with zero mean and unit variance, satisfying the ensemble property ($\langle \xi_n\xi_m = \delta_{nm} \rangle$).

To generate non-Markovian trajectories, the correlation matrix $C_{ij}$ must be decomposed into a lower-upper triangular matrix ${\bf{L}}$, such that $C={\bf{L}}{\bf{L}}^T$. Since the correlation matrix is real, symmetric, and positive definite, the decomposition yields a unique triangular matrix solution, which can be efficiently computed using the well-known Cholesky algorithm.

It is important to note that the correlation matrix is an ill-posed matrix with a near-zero determinant, requiring careful numerical treatment to ensure computational stability. Once properly handled, the ensemble of trajectories can be generated. The density contrast at scale $R_i$ is given by:

\be
\delta_i = \sum_{j} {\bf{L}}_{ij} \xi_j,
\ee
where $\xi_j$ is random number with Gaussian distribution. In this case, $\delta_i$ will have the correct correlation between heights which is given by
\be
\langle \delta_i \delta_j \rangle = \sum_{m,n} L_{im} L_{jn} \langle \xi_m \xi_n \rangle = {\bf{L}}{\bf{L}}^T = C.
\ee
For a more detailed discussion on non-Markov trajectories and the Cholesky decomposition algorithm, refer to \cite{Nikakhtar:2018qqg,kameli:2020mod}.  

Figure \ref{fig:nonmarkov} presents a comparison between Markov and non-Markov trajectories. The non-Markov trajectories are generated using the aforementioned Cholesky algorithm, applied to a Gaussian window function $\tilde{W}(x) = \exp[-x^2/2]$. Due to technical constraints, this window function has been consistently employed throughout our entire calculations.

\section{EST-model comparison} \label{app:II}

\begin{figure}
	\includegraphics[width=\columnwidth]{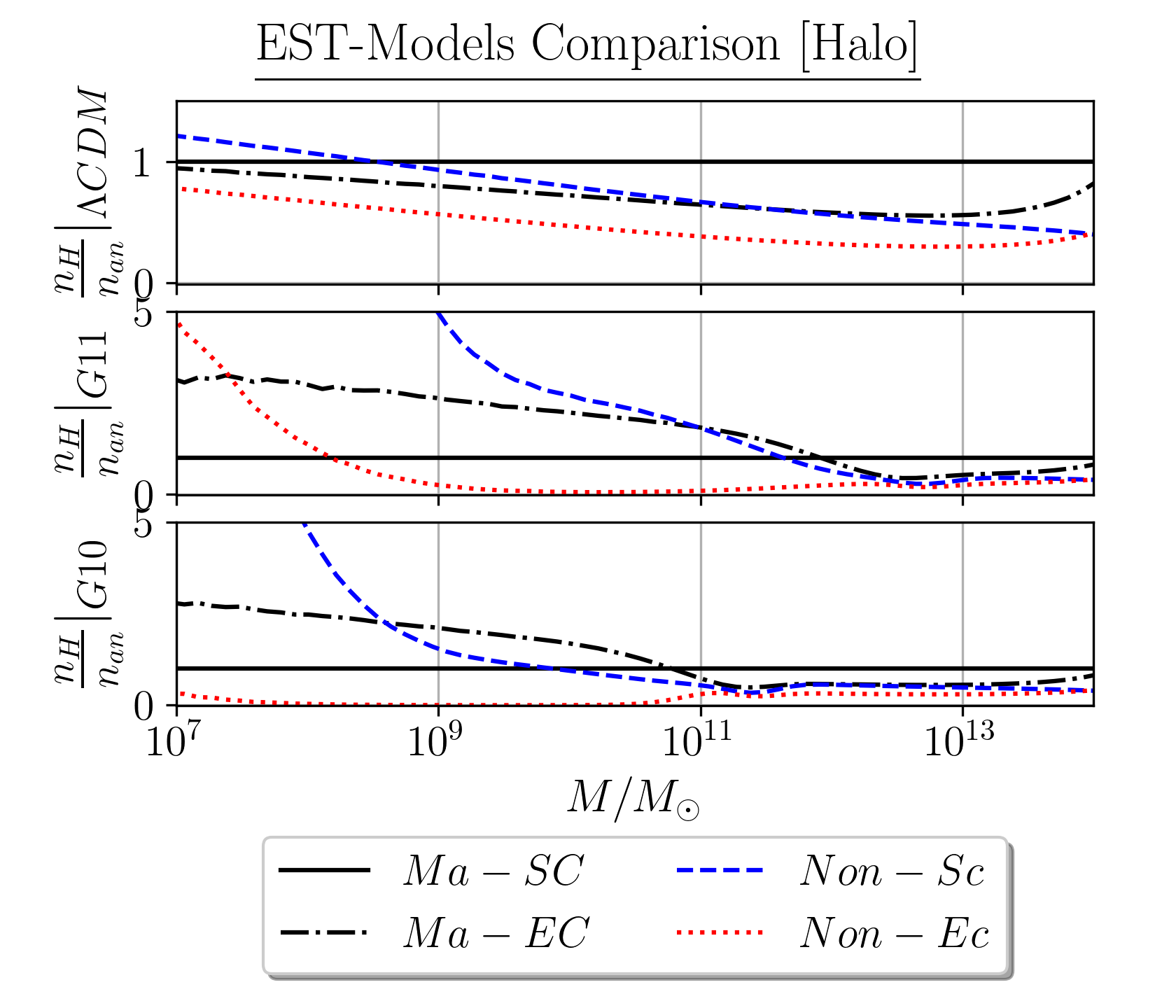}
	\caption{The comparison of EST models for the number density of halos versus halo mass is presented. The ratios of various EST models (H) to the Markov-SC model as the baseline (an), including Markov (Ma) and Non-Markov (Non) cases, as well as Spherical (SC) and Ellipsoidal (EC) collapse models, are plotted. The three panels correspond to the standard $\Lambda$CDM model, G11, and G10 models, respectively.} \label{fig:HaloC}
\end{figure}

\begin{figure}
	\includegraphics[width=\columnwidth]{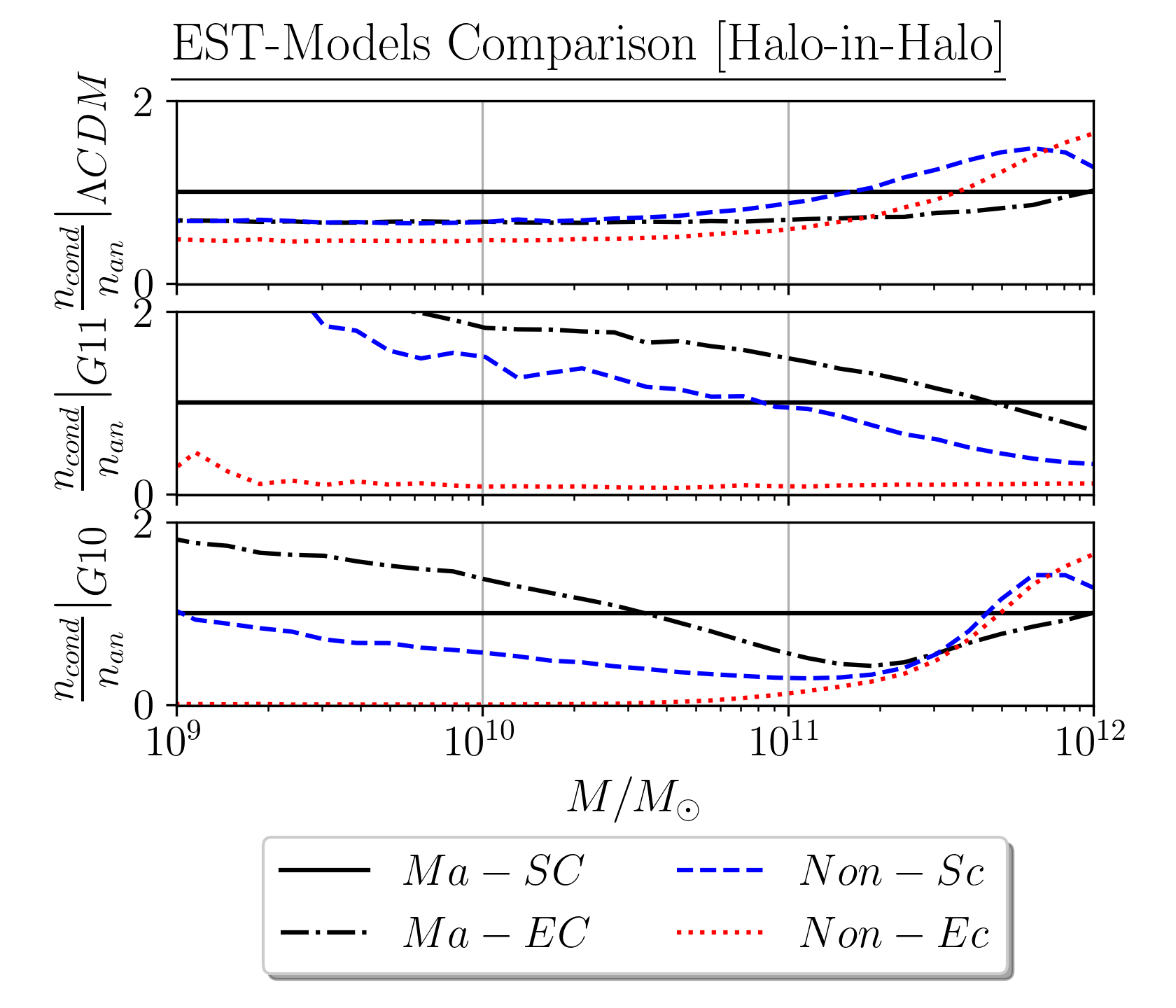}
	\caption{The comparison of EST models for the number density of Halo-in-Halos versus progenitor halo mass is presented. The conditional number density of progenitor halos at redshifts $z_2=2$, which eventually form a host halo with $M=10^{13} M_\odot$ at $z_0=0$, is presented. The ratios of various EST models (Cond) to the Markov-SC model as the baseline (an), including Markov (Ma) and Non-Markov (Non) cases, as well as Spherical (SC) and Ellipsoidal (EC) collapse models, are plotted. The three panels correspond to the standard $\Lambda$CDM model, G11, and G10 models, respectively.} \label{fig:HiHC}
\end{figure}

\begin{figure}
	\includegraphics[width=\columnwidth]{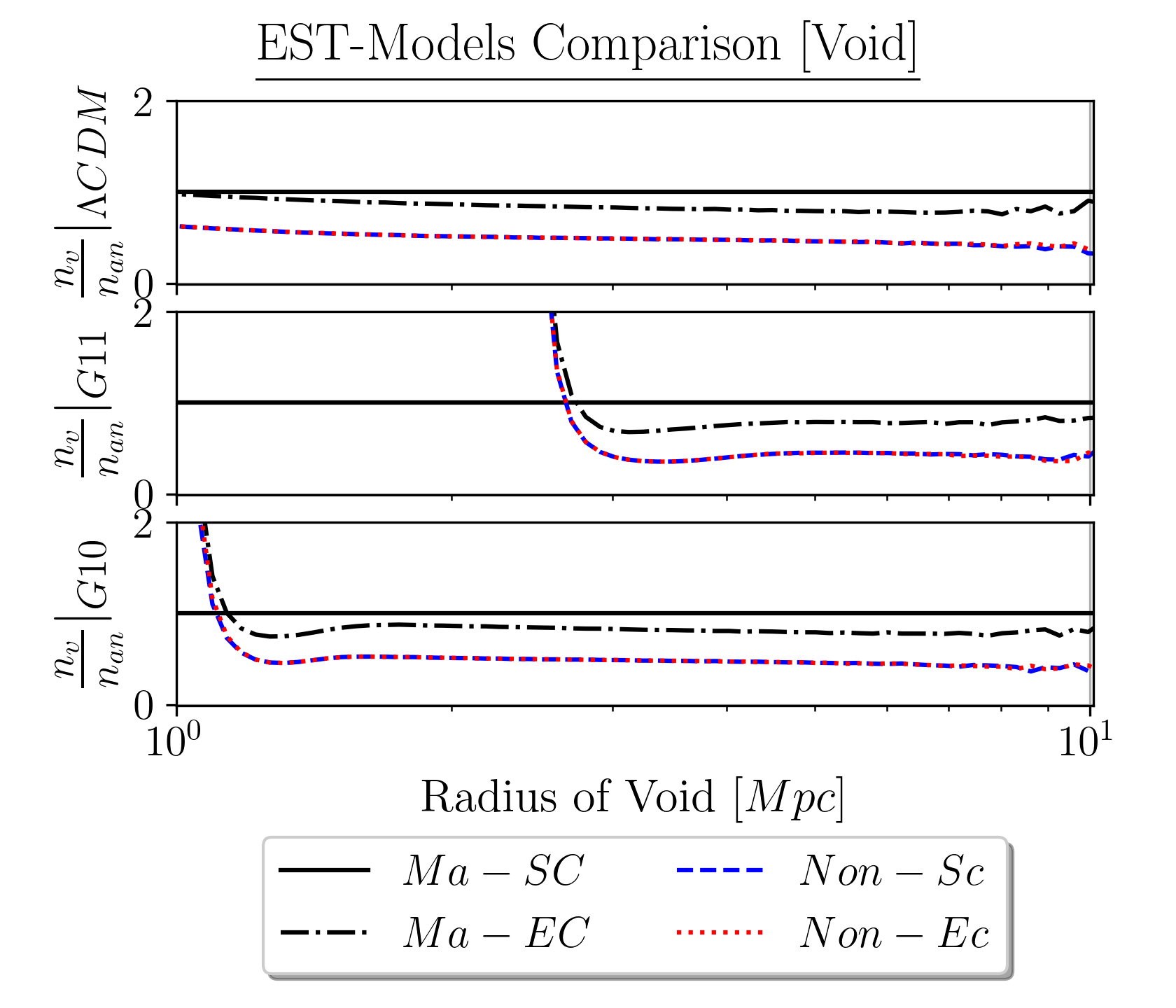}
	\caption{The comparison of EST models for the number density of voids versus Eulerian radius is presented. The ratios of various EST models (V) to the Markov-SC model as the baseline (an), including Markov (Ma) and Non-Markov (Non) cases, as well as Spherical (SC) and Ellipsoidal (EC) collapse models, are plotted. The three panels correspond to the standard $\Lambda$CDM model, G11, and G10 models, respectively..} \label{fig:VoidC}
\end{figure}

\begin{figure}
	\includegraphics[width=\columnwidth]{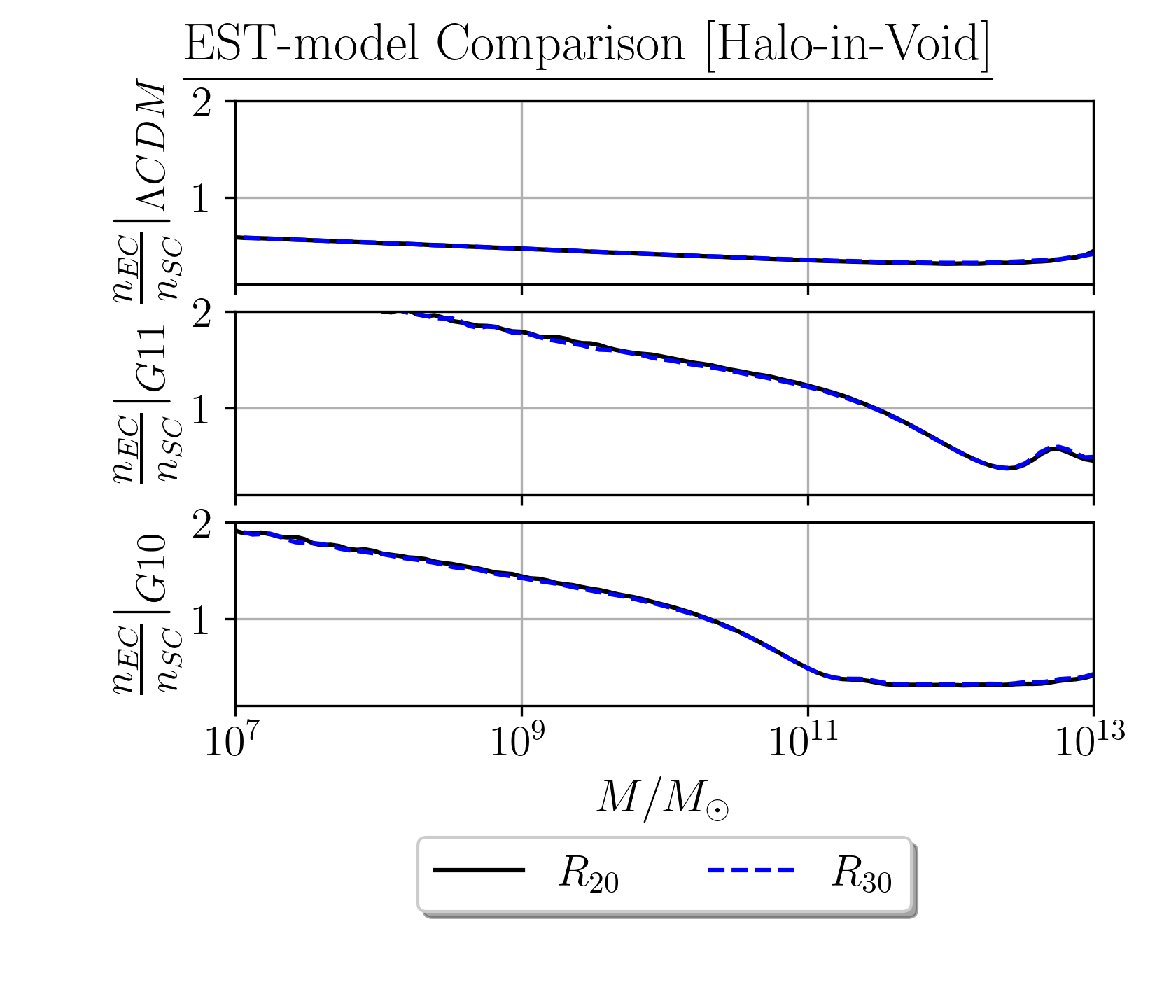}
	\caption{The comparison of EST models for the number density of Halo-in-Voids versus embedded halo mass is presented. The ratios of EC-model to SC-model for voids with radius  $R_{20}=20 ~\rm{Mpc}$ and $R_{30}=30 ~ \rm{Mpc}$, are plotted. The three panels correspond to the standard $\Lambda$CDM model, G11, and G10 models, respectively.} \label{fig:HiVC}
\end{figure}

In this appendix, we analyze and compare the outcomes of different EST-models for halo number density, halo-in-halo, void, and halo-in-void statistics. As discussed previously, the various EST-models exhibit no fundamental or significant influence on the findings, particularly on the ratios of the modified models to the standard cosmological model. Instead, they primarily serve as corrections to the analytical results of the Markov-SC model. Here, we provide a more detailed examination of these corrections. For each specific statistic (e.g., halo number density), we present three subplots corresponding to the standard $\Lambda$CDM model, the G11 model, and the G10 model. Each subplot displays the ratio of the various applied EST-models to the baseline Markov-SC model (indicated by black lines with  the value of 1.0), with each line representing the ratio for a distinct EST-model within a particular cosmological model.

Fig. \ref{fig:HaloC} illustrates the ratios of different EST-models for each halo number density. For the standard model, the non-Markov models predict a higher number density of small halos and a reduction in massive halos. This behavior arises from the smoothed nature of non-Markov trajectories, which results in a slower rate of increase in the density contrast ($\delta_L$). Consequently, the likelihood of first up-crossing at small variance (large mass) decreases compared to Markov trajectories, while crossings at larger variance (smaller masses) become more frequent. This phenomenon is even stronger in the G11 and G10 models, suggesting that modified models may produce more small halos relative to the analytical model. The Markov-EC model yields lower number density values due to the variance-dependent moving barrier. As shown in Fig. \ref{fig:traj}, this moving barrier exceeds the constant SC barrier except at large scales (small S). Finally, the non-Markov EC model incorporates a combination of all these effects. Fig. \ref{fig:HiHC} presents analogous results for the ratio of halo-in-halo number densities to the analytical baseline for different EST-models. In this case, a greater abundance of massive progenitors is observed near the host halo's scale. In the G11 model this feature will be seen for halos with masses exceeding $M\simeq 10^{13} \, M_\odot$.

Similarly, Fig. \ref{fig:VoidC} compares the EST-models for void statistics. While these results do not correspond to void radii of observational interest, smaller ratios are observed for other EST-models relative to the analytical baseline. The Markov-EC model exhibits slightly lower values, likely due to the exclusion of trajectories crossing the collapse barrier at very small variance. At such small variance, the ellipsoidal barrier is smaller compared to the SC barrier. Non-Markov models further suppress the statistics due to their smoother trajectories. In Fig. \ref{fig:HiVC}, we present only the ratio of EC to SC models for different void radii. As noted previously in the results, non-Markov outcomes could not be obtained. The EC model values are generally reduced compared to the SC model, particularly at large mass scales for the standard cosmological model. In halo-in-void statistics, the second crossing of the collapse barrier occurs at bigger variance, resulting in smaller ratios at large mass scales. The modified models exhibit larger values at small mass scales, likely due to a slower rate of decline in the ratios for these models.

In conclusion, we confirm that the various EST models exert no significant impact on the overall results. Presenting the ratio outcomes for different cosmological models ensures consistency in our findings and reveals broadly similar results for all EST-models. The primary focus of this manuscript is the comparison of different cosmological models, while the variations among EST-models playing a relatively minor role.

\end{twocolumn}

\bsp	
\label{lastpage}
\end{document}